\documentclass{emulateapj}
\usepackage{natbib}
\usepackage{bm}
\usepackage{mathrsfs,amsmath}
\usepackage{graphicx}
\usepackage{epstopdf}
\usepackage[english]{babel}
\usepackage[colorlinks,linkcolor={blue},citecolor={blue},urlcolor={blue}]{hyperref}
\newcommand{\mpchi}{\,h^{-1}{\rm {Mpc}}}

\newcommand{\msun}{M_{\sun}}
\newcommand{\wrp}{w_{\rm p}}

\newcommand{\wprp}{w_{\rm p}(r_{\rm p})}

\shorttitle{Stellar mass functions of galaxies at $0.1<z<0.8$}
\shortauthors{H. Guo et al.}
\begin{document}
\title{The Incomplete Conditional Stellar Mass Function: Unveiling the Stellar Mass Functions of Galaxies at $0.1 < Z < 0.8$ from BOSS Observations}

\author{Hong Guo\altaffilmark{1}, Xiaohu Yang\altaffilmark{2,3}, Yi
  Lu\altaffilmark{1}}

\altaffiltext{1}{Key Laboratory for Research in Galaxies and
  Cosmology, Shanghai Astronomical Observatory, Shanghai 200030,
  China; guohong@shao.ac.cn}

\altaffiltext{2}{Department of Astronomy, Shanghai Key Laboratory for
Particle Physics and Cosmology, Shanghai Jiao Tong University,
Shanghai 200240, China}

\altaffiltext{3}{IFSA Collaborative Innovation Center, and Tsung-Dao Lee
Institute, Shanghai Jiao Tong University, Shanghai 200240, China}

\begin{abstract}
  We propose a novel method to constrain the missing fraction of
  galaxies using galaxy clustering measurements in the galaxy
  conditional stellar mass function (CSMF) framework, which is
  applicable to surveys that suffer significantly from sample
  selection effects. The clustering measurements, which are not
  sensitive to the random sampling (missing fraction) of galaxies, are
  widely used to constrain the stellar-halo mass relation (SHMR). By
  incorporating a missing fraction (incompleteness) component into the
  CSMF model (ICSMF), we use the incomplete stellar mass function and
  galaxy clustering to simultaneously constrain the missing fractions
  and the SHMRs.  Tests based on mock galaxy catalogs with a few
  typical missing fraction models show that this method can accurately
  recover the missing fraction and the galaxy SHMR, and hence provides
  us reliable measurements of the galaxy stellar mass functions. We
  then apply it to the Baryon Oscillation Spectroscopic Survey (BOSS)
  over the redshift range of $0.1<z<0.8$ for galaxies of
  $M_*>10^{11}\msun$. We find the sample completeness for BOSS is over
  80\% at $z<0.6$, but decreases at higher redshifts to about
  30\%. After taking these completeness factors into account, we
  provide accurate measurements of the stellar
  mass functions for galaxies with $10^{11}\msun<M_*<10^{12}\msun$, as
  well as the SHMRs, over the redshift range $0.1<z<0.8$ in this
  largest galaxy redshift survey.
\end{abstract}
	
\keywords{cosmology: observations --- cosmology: theory --- galaxies:
  distances and redshifts --- galaxies: halos --- galaxies: statistics
  --- large-scale structure of universe}
	
\section{Introduction}

The connection between the galaxy properties and those of the dark
matter has been investigated in depth in the past decades for the
local and high-redshift galaxies \citep[see e.g.,][]{Norberg01,Zehavi02,Zehavi05, Zehavi11,Yang03,Yang04,Yang07,
Zheng07,Moster10,Moster13, Coupon12, Leauthaud12, Guo14, McCracken15}. The galaxy stellar-halo mass relation (SHMR),
in particular, provides important constraints to the galaxy formation
and evolution models \citep[see e.g.,][]{Yang09, Yang12, Moster10, Moster13, Behroozi13,Beutler13,Reddick13, Lin16, Zu15, Zu16,Wang18}, as it probes the
joint-evolution of the galaxy and halo mass growth histories and is
directly related to the cosmic star formation histories \citep[see
e.g.,][]{Behroozi13,Yang13}.

There are multiple ways of determining the halo masses for galaxies of
different stellar masses. The common methods include the halo
occupation distribution (HOD) \citep[see e.g.,][]{Jing98,Peacock00,Seljak00,Scoccimarro01,Berlind02,Berlind03,Zehavi05,
  Zheng05,Zheng07,Zheng09,Guo14, Guo15,McCracken15} and conditional stellar mass
function (CSMF) \citep[see e.g.,][]{Yang09,Yang12,More12,Cacciato13,More13,Reddick13,Bosch08,Bosch13,Yang17b} modeling of the galaxy clustering measurements, the subhalo abundance matching models
\citep[see e.g.,][]{Rodriguez-Puebla12, Rodriguez-Puebla17,
  Behroozi13, Moster13, Guo16}, and the direct weak gravitational
lensing measurements \citep[see e.g.,][]{Mandelbaum06, Miyatake15,
  Zu15}. The central galaxy SHMR has been extensively studied with
these methods for galaxy samples at different redshifts, and is found
to follow a broken power law relation with a steep slope for low mass
halos of $M<10^{12}\msun$ and becoming much flatter for more massive
halos \citep{Yang09, Yang12, Moster10, Moster13, Wang10}.  Within all
these probes, a crucial measurement is the galaxy stellar mass function (SMF).

Thanks to the large-scale galaxy redshift surveys, e.g., the 2dF
Galaxy Redshift Survey \citep[2dFGRS;][]{Colless99} and the Sloan
Digital Sky Survey \citep[SDSS;][]{York00}, at the low redshifts of
$z<0.2$, we are able to accurately measure the galaxy stellar mass
function \citep[see e.g.,][]{Bell03,Li09, Bernardi10,Bernardi13}
and the SHMR \citep[see e.g.,][]{Yang09, Yang12, Wang07,Moster10,
  Moster13,Behroozi13,Rodriguez-Puebla17}. At higher redshifts, the
measurements of the galaxy SMFs in literature have much larger
uncertainties compared to the low-redshift ones, because the galaxy
SMFs at high redshifts are mostly derived from deep photometric
surveys covering small sky area, where the sample variance effect
would dominate the error budget \citep{Davidzon13,Davidzon17,Ilbert13,Maraston13,Muzzin13,Moustakas13,Tomczak14,Santini15}. For example, as summarized in Table 1
of \cite{Rodriguez-Puebla17} (see also Table 3 of
\citealt{Behroozi13}), the galaxy SMF at $0.2<z<1$ is measured with
the largest volume in \cite{Moustakas13} using the PRism MUlti-object
Survey \citep[PRIMUS;][]{Coil11} covering an area of 9~deg$^2$.

As constraints to the galaxy SHMR are generally obtained from fitting
the galaxy SMFs with or without the spatial clustering measurements,
models of the SHMR for higher-redshift galaxies thus have
significant differences among different studies, although they agree
with each other within errors for low-redshift galaxies \citep[see
e.g.,][]{Shankar14}. The difference is more significant for massive
galaxies of $M_*>10^{11}\msun$. For example, the SHMR model of
\cite{Yang12} would predict a high-mass end slope of
$M_*\propto M^{0.375}$ for galaxies at $z=0.5$, while of the model of
\cite{Behroozi13} would have $M_*\propto M^{0.265}$. However, the
total number densities of galaxies with $M_*>10^{11}\msun$ for these
two models only differ by about 20\%, as the halo mass function is
decreasing very fast toward the high mass end. Therefore,
discriminating the different SHMR models would require accurate
measurements of the galaxy SMF at the massive end, which can only be
achieved with wide-area galaxy surveys. Moreover, accurate clustering
measurements to constrain the SHMR models would also require a large
sample volume.

Currently, the largest galaxy redshift survey at $0.2<z<0.8$ is the
SDSS-III Baryon Oscillation Spectroscopic Survey
\citep[BOSS;][]{Dawson13}. The latest data release 12 of the BOSS
galaxy sample covers an area of 10,252~deg$^2$ \citep{Reid16}, which
is about 1100 times larger than that of the PRIMUS. Therefore,
the dominating errors on the BOSS galaxy SMF measurements come from
the systematic errors on the galaxy stellar mass measurements, rather
than the sample variance effect, which only has a minor
contribution. The BOSS galaxy sample would potentially provide the
most accurate galaxy SMF and SHMR measurements at these intermediate
redshifts. However, the main science drive for BOSS is to
measure the baryonic acoustic oscillation (BAO) signals in order to
constrain the cosmological parameters \citep[see e.g.,][]{Alam17}, so
the galaxy targeting strategy is to select intermediate-redshift
luminous galaxies that cover a large enough volume. Due to the
complicated selection criteria of both the apparent magnitude and
color, the resulting galaxy sample is, however, neither volume-limited
nor stellar mass complete. Thus, the measured galaxy SMFs in BOSS
cannot be directly compared with those from other studies of
volume-limited samples. Furthermore, the BOSS galaxy sample is also not a homogeneous sample of luminous red galaxies at the intermediate redshifts, but also purposely includes a significant fraction of blue galaxies \citep{Maraston13}.

Efforts have been made to estimate the stellar mass completeness for
the BOSS galaxies, using either galaxies selected with wider color
cuts \citep{Tinker17} or deeper imaging observations of those massive
galaxies in the Stripe 82 region \citep{Leauthaud16,Saito16}, with the
conclusions that the high redshift BOSS galaxies are significantly
affected by the mass incompleteness. In fact, the mass incompleteness
caused by the complicated target selection is not just the concern of
BOSS. It has become a common issue for many large-scale
galaxy redshift surveys targeting at high-redshift objects. For
example, the on-going SDSS-IV extended Baryon Oscillation
Spectroscopic Survey \citep[eBOSS;][]{Dawson16}, targeting at the
luminous red galaxies \citep[LRG;][]{Prakash16} and emission line
galaxies \citep[ELG;][]{Raichoor17} in the redshift range of
$0.7<z<1.1$ adopts various apparent magnitude and color cuts to select
objects for further spectroscopic observations. Similar issues also
happen for many of the next-generation galaxy redshift surveys, such
as the Dark Energy Spectroscopic Instrument \citep[DESI;][]{DESI16}
and Prime Focus Spectrograph \citep[PFS;][]{Takada14}.

Naturally, it is more reliable to constrain the missing fraction of galaxies caused by the complicated target selections without resort to external measurements from other surveys. \cite{Montero16} presented a forward-modeling technique to quantify the completeness of BOSS galaxies in the red sequence at $z\sim0.55$. By matching the observed color-magnitude distributions with reasonable analytical parametric models convolved with the photometric errors and selection effects, it is possible to derive the intrinsic color-magnitude distribution and therefore estimate the completeness as a function of magnitude. In this paper, we propose a novel method to constrain the completeness by making use of a clustering measurement property---the galaxy clustering
is insensitive to the random sampling (missing fraction) of
galaxies.  On the other hand, within the CSMF framework, accurate clustering measurements can be used
constrain the SHMR. We can therefore incorporating a missing fraction
component in the SHMR, so that both the SHMR and the missing fraction
of galaxies can be simultaneously constrained. We used mock galaxy
catalogs with a few typical missing fraction models to demonstrate the
reliability of such a method and then apply it to the BOSS galaxy
sample to provide so far the most accurate measurements of the galaxy SMFs and SHMRs at massive end
in the redshift range of $0.1<z<0.8$.
	
The structure of the paper is constructed as follows. In
\S\ref{sec:data}, we describe the galaxy samples and the simulation
used in the modeling. We introduce our modeling method in
\S\ref{sec:method} and test it in \S\ref{sec:test}. We present the
modeling results for the BOSS galaxies in \S\ref{sec:results} and
discuss our models in \S\ref{sec:discussion}. We summarize our results
in \S\ref{sec:conclusion}.  Throughout this paper, we assume a
spatially flat $\Lambda$ cold dark matter cosmology, with
$\Omega_{\rm m}=0.307$, $h=0.678$, $\Omega_{\rm b}=0.048$ and
$\sigma_8=0.823$, consistent with the constraints from Planck
\citep{PlanckCollaboration14} and with the parameters used in the
simulation adopted for our modeling (see \S\ref{sec:data}). For the
stellar mass estimates, we assume a universal \cite{Chabrier03}
initial mass function (IMF), the stellar population synthesis model of
\cite{Bruzual03} and the time-dependent dust attenuation model of
\cite{Charlot00}. All masses are in units of $\msun$.
	
\section{Data} \label{sec:data}

\subsection{BOSS Galaxy Sample}

We use the Data Release 12 of the BOSS galaxy sample, with redshifts
for 1,372,737 galaxies over an area of 10,252~deg$^2$
\citep{Reid16}. The detailed descriptions of the survey can be found
in \cite{Eisenstein11} and \cite{Dawson13}. The BOSS galaxy sample is
formally divided into two subsamples with different target selections
focusing on galaxies at low and high redshifts. The target selections
are based on following set of combinations of model magnitudes,
\begin{eqnarray}
  c_{||} &= &  0.7(g-r) + 1.2(r-i - 0.18), \\
  c_{\perp} &= & (r-i) - (g-r)/4 - 0.18, \\
  d_{\perp} &= & (r-i) - (g-r)/8,
\end{eqnarray}
where the $g-r$ and $r-i$ colors are based on the \texttt{model}
magnitudes.

The low redshift sample, denoted as LOWZ, is an extension to the
SDSS-II LRG sample \citep{Eisenstein01} to fainter magnitudes in
$0.15<z<0.43$ , with the selection cuts of 
\begin{eqnarray}
|c_{\perp}|<0.2, \\
r<13.6+c_{||}/0.3, \\
16<r<19.6,
\end{eqnarray}
where $r$ is based on the \texttt{cmodel} magnitude. As clearly shown in Figure~11 of \cite{Eisenstein01}, most of the galaxies selected with the above cuts in LOWZ would be LRGs.

The higher redshift sample, denoted as
CMASS, targets galaxies that roughly follow a constant stellar mass
cut in the redshift range of $0.43<z<0.7$ \citep{Maraston13}. The
corresponding selection cuts are 
\begin{eqnarray}
|d_{\perp}|>0.55, \\
i<19.86+1.6(d_{\perp}-0.8), \\
17.5<i<19.9,
\end{eqnarray}
where the magnitude $i$ is the $i$-band \texttt{cmodel} magnitude.

To increase the stellar mass completeness at different redshifts, we
use the combined sample of LOWZ and CMASS. The redshift and angular
distribution of the combined BOSS galaxy sample are presented in
\cite{Reid16} (their Figures 8 and 11). The sky coverage of the LOWZ
sample in the northern galactic cap (NGC) is slightly smaller than
that of CMASS, due to the removal of data observed in the first nine
months that have the incorrect star-galaxy separation cut applied. At
$z<0.4$, as there is only a small fraction of CMASS galaxies in this
redshift range, we include all LOWZ galaxies and those CMASS galaxies
that fall into the LOWZ geometry mask (covering about
$9000$~deg$^2$). At $z>0.4$, the different angular distributions of
LOWZ and CMASS are taken into account by combining the random catalogs
for the two samples when measuring the spatial clustering.

As we have a combined sample covering a large redshift range of
$0.1<z<0.8$, we divide the galaxies into different redshift intervals
with a bin size of $\Delta z=0.1$ to study the evolution of these
massive galaxies. In total, we have seven redshift bins, with the
detailed information displayed in Table~\ref{tab:data}, where the
total number of galaxies, $N_{\rm tot}$, the average sample stellar
mass with the corresponding scatter, $\langle\log M_*\rangle$, and the
mean number density, $\bar{n}_{\rm g}$, are shown for each sample. As
the galaxy SMF and SHMR have been studied extensively in $0.1<z<0.2$
with the SDSS DR7 Main galaxy sample, in this paper we focus on the
measurements in $0.2<z<0.8$, and use the low-redshift measurements as
a consistency check with the literature.

\begin{table}
	\caption{Samples of Different Redshift Bins} \label{tab:data}
	\centering
	\begin{tabular}{lccc}
          \hline
          redshift range  & $N_{\rm tot}$ & $\langle\log 
                                            (M_*/\msun)\rangle$
          &$\bar{n}_{\rm g}/(h^3\, \rm{Mpc}^{-3})$ \\
          \hline	
          $0.1<z<0.2$  & 84404 & $11.02\pm0.12$ & $5.71\times10^{-4}$\\			
          $0.2<z<0.3$  & 117822 & $11.10\pm0.12$ & $3.25\times10^{-4}$\\	
          $0.3<z<0.4$  & 179726 & $11.30\pm0.12$ & $2.84\times10^{-4}$\\	
          $0.4<z<0.5$  & 294291 & $11.31\pm0.14$ & $2.80\times10^{-4}$\\	
          $0.5<z<0.6$  & 398565 & $11.28\pm0.15$ & $2.85\times10^{-4}$\\		
          $0.6<z<0.7$  & 172251 & $11.37\pm0.16$ & $0.99\times10^{-4}$\\	
          $0.7<z<0.8$  & 30954 & $11.64\pm0.69$ &  $0.15\times10^{-4}$\\	                			
          \hline
	\end{tabular}
	 

\end{table}
\begin{figure}
	\centering
	\includegraphics[width=0.45\textwidth]{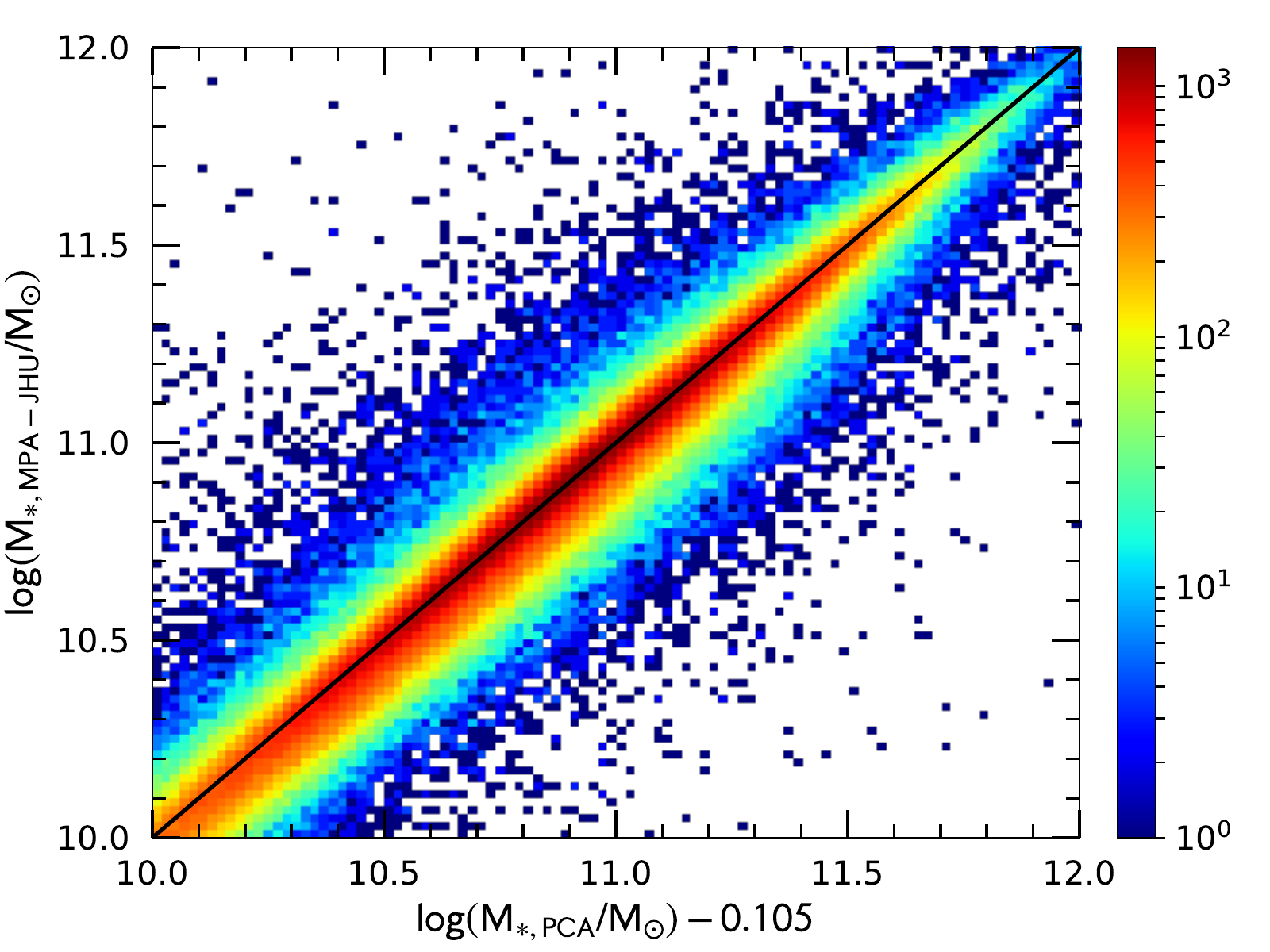}
	\caption{Comparisons between the stellar masses derived from
          the PCA method of \cite{Chen12} and those from the MPA-JHU
          catalog. The PCA stellar masses are systematically
          overestimated by 0.105~dex. The color scale shows the
          density distribution.  }
	\label{fig:mstar}
\end{figure}

The galaxy stellar mass used in this paper is estimated in
\cite{Chen12} by fitting the galaxy spectra over the rest-frame
wavelength range of 3700--5500$\rm{\AA}$ with the principal component
analysis (PCA) method. We use the stellar mass obtained by applying
the stellar population synthesis (SPS) models of \cite{Maraston11}
with an IMF of \cite{Kroupa01} and a dust attenuation model of
\cite{Charlot00}. The total galaxy stellar mass is obtained by
applying the mass-to-light ratio within the fiber aperture to the
whole galaxy. We refer the readers to \cite{Chen12} for details.

At low redshifts, the BOSS sample (especially LOWZ) has a significant
overlap with the SDSS DR7 Main galaxy sample \citep{Abazajian09},
where the galaxy stellar masses have been derived in the MPA-JHU
catalog\footnote{https://wwwmpa.mpa-garching.mpg.de/SDSS/DR7/}
following the method of \cite{Kauffmann03} by applying the SPS model
of \cite{Bruzual03}. In order to compare with the literature for
low-redshift measurements of the galaxy SHMR and SMFs, we cross-match
the two galaxy samples and show the comparisons of the stellar masses
in the different catalogs in Figure~\ref{fig:mstar}. The stellar
masses with the PCA method are systematically overestimated by
0.105~dex compared to the MPA-JHU stellar masses, due to the truncated
star formation histories and a smaller fraction of galaxies with
recent bursts \citep[see Figures~12 and 13 of][and discussions therein]{Chen12}. We also apply
the standard correction of a constant shift of -0.05~dex
\citep{Bernardi10} to convert from a \cite{Kroupa01} IMF to that of
\cite{Chabrier03}. Therefore, in this paper we reduce the galaxy
stellar masses in \cite{Chen12} by 0.155~dex to be consistent with
literature.

\begin{figure}
	\centering
	\includegraphics[width=0.45\textwidth]{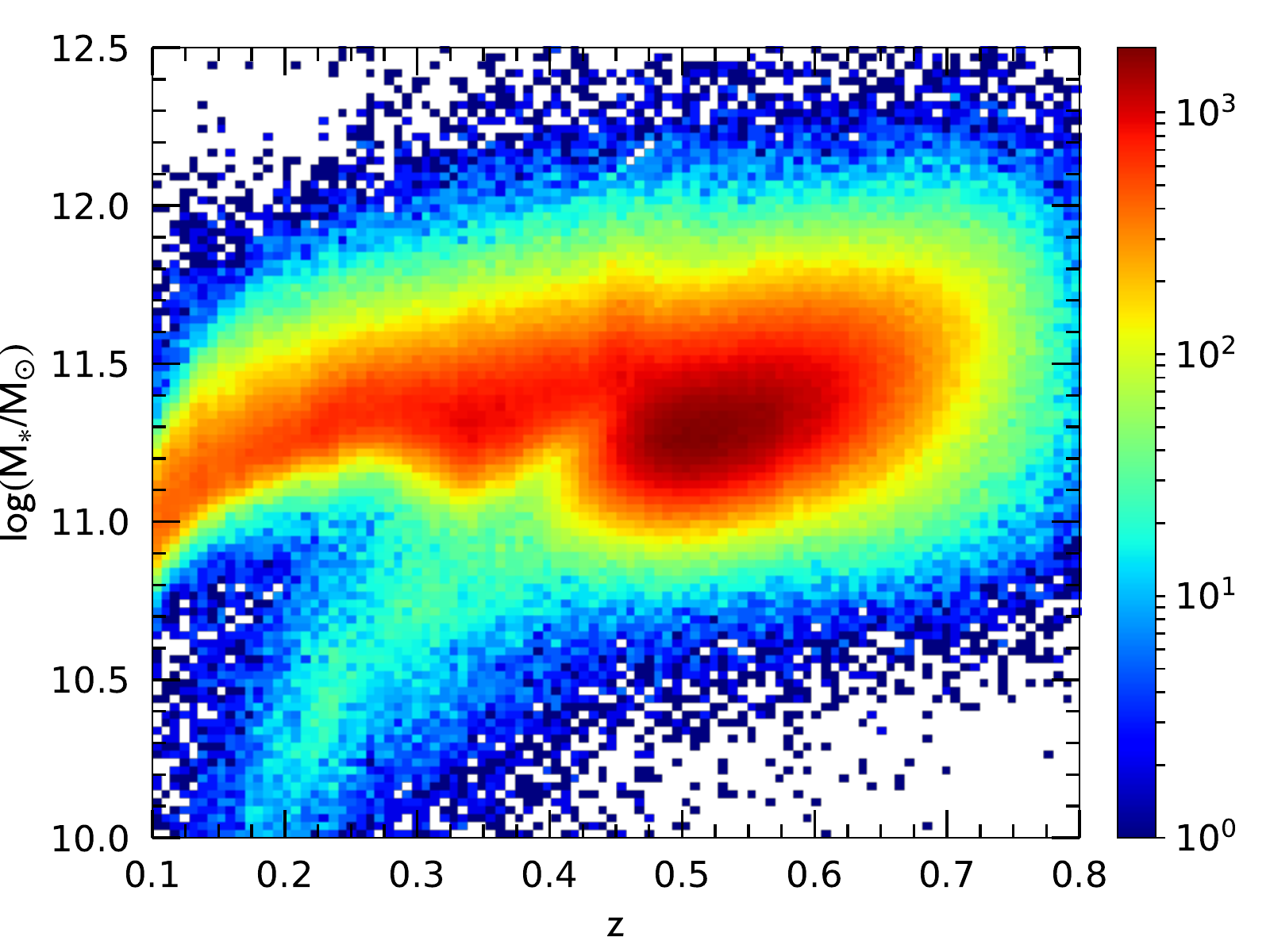}
	\caption{Stellar mass distribution as a function of redshift
          $z$ for the combined LOWZ+CMASS sample. The color scale
          shows the density distribution.}
	\label{fig:msz}
\end{figure}

We show in Figure~\ref{fig:msz} the stellar mass distribution as a
function of redshift for the combined sample. The majority of the BOSS
galaxies are more massive than $10^{11}\msun$. We can therefore
accurately measure the massive end of the galaxy SMF and model the
corresponding SHMR. As the redshift increases, the average galaxy
stellar mass moves toward slightly more massive end due to the flux
limit of the sample selection criteria. In order to study the evolution of galaxy SMF in a broad redshift range, we only focus on the massive galaxies with $M_*>10^{11}\msun$ in this paper. 
	
As will be demonstrated in the following sections, although red and blue galaxies could have different selection functions, the derived total galaxy SMFs and the SHMRs using the whole BOSS sample are not affected, even if we only select homogeneous red galaxy samples in our measurements. Because the red galaxies dominate the contribution to the SMF for $M_*>10^{11}\msun$ \citep[see e.g., Figure 10 of][]{Moustakas13}, the small fraction of blue galaxies in BOSS would not have any significant effect on the SHMR for the whole galaxy sample, even though red and blue galaxies tend to have slightly different SHMRs.

\subsection{Dark Matter Simulation}

In order to evaluate our method in constraining the missing fraction
of galaxies, we use the dark matter halos extracted from the
BigMultidark Planck simulation
\citep[BigMDPL\footnote{https://www.cosmosim.org/cms/simulations/bigmdpl/};][]{Klypin16},
with the cosmological parameters of $\Omega_{\rm m}=0.307$,
$\Omega_{\rm b}=0.048$, $h=0.678$, $n_{\rm s}=0.96$ and
$\sigma_8=0.823$. The simulation has a box size of
$2.5\,h^{-1}{\rm{Gpc}}$ and a mass resolution of
$2.4\times10^{10}h^{-1}M_\odot$. It has been used to construct mock
galaxy catalogs for the CMASS sample \citep{Rodriguez-Torres16}. The
dark matter halos and subhalos in the simulation are identified with
the \texttt{ROCKSTAR} phase-space halo finder \citep{Behroozi13b}.

We use seven different redshift outputs of $z=0.152$, 0.246, 0.347,
0.453, 0.547, 0.655, and 0.759 from BigMDPL, roughly corresponding to
the median redshifts of the seven BOSS galaxy samples in the different
redshift bins. The galaxy clustering and stellar mass function
measurements from the different redshift bins are modeled separately
using the corresponding halo catalogs in order to constrain their
evolution.

\section{Method} \label{sec:method}

In this paper, we incorporate a incompleteness component into the CSMF
(hereafter, ICSMF) framework \citep{Yang12} to predict the accurate
galaxy total SMF and SHMR from incomplete galaxy samples as those in
BOSS.

\subsection{Theoretical Models}

In this subsection, we provide the details of our ICSMF model, as well
as the model predictions of the stellar mass functions and clustering
properties of galaxies (i.e., the projected two pointed correlation
functions).

\subsubsection{Incomplete Conditional Stellar Mass Function}

Following \cite{Yang12}, we assume that the CSMF of central galaxies,
i.e., the average number of central galaxies with stellar mass $M_*$
hosted by halos of mass $M$, can be characterized by a log-normal
distribution function,
\citep{Yang09},
\begin{equation}
  \Phi_{\rm c}(M_*|M)=\frac{1}{\sqrt{2\pi}\sigma_{\rm c}}\exp
  \left[-\frac{(\log M_*- \log\langle M_*|M\rangle)^2} {2 \sigma_{\rm c}^2}\right] \label{eq:csmf}
\end{equation}
where $\sigma_{\rm c}$ is the scatter of stellar mass distribution in
a given halo with mass $M$. The scatter is not well constrained by the
SMF only \citep[see e.g.,][]{Moster10}. It is generally assumed to be
a constant value of around $0.173$ (it is assumed to be 0.15 in
\citealt{Moster13}) for low redshifts of $z<0.9$
\citep{Yang12}. Recent measurements of \cite{Tinker17} suggest that
the scatter $\sigma_{\rm c}$ can be reasonably constrained with the
galaxy clustering measurements, where they estimated $\sigma_{\rm c}$
for the BOSS CMASS galaxies to be around 0.18, consistent with that of
\cite{Yang12}. In this paper, we simply set the intrinsic scatter
$\sigma_{\rm c}$ to be a constant of $0.173$ in the redshift range of
$0.1<z<0.8$. We will discuss the effect of different $\sigma_{\rm c}$
values in Section~\ref{sec:discussion}.

The function $\langle M_*|M\rangle$ is the average central galaxy
stellar mass in halos of mass $M$. We adopt a broken power law
relation as in \cite{Yang12} \citep[see also][]{Moster10,Moster13},
\begin{equation}
 \langle M_*|M\rangle = M_{*,0}\frac{(M/M_1)^{
  		\alpha+\beta}}{(1+M/M_1)^{\beta}} \label{eq:smhm}
\end{equation}
where $M_{*,0}$, $M_1$, $\alpha$, and $\beta$ are the four free parameters in
this relation. The values of $\alpha+\beta$ and $\alpha$ are the low and
high-mass end slopes of the SHMR, respectively.

To properly estimate the missing fraction of galaxies in the BOSS
observation, \cite{Leauthaud16} derived the stellar mass completeness
by comparing the observed (incomplete) BOSS galaxy SMFs with the total
SMFs obtained by combining the measurements of PRIMUS and deeper
imaging of the stripe 82 region \citep{Bundy15}. They found that the
stellar mass completeness of BOSS galaxies at different redshifts can
be well described by the following functional form,
\begin{equation}
  c(M_*)=\frac{f}{2}\left[1+{\rm erf}\left(\frac{\log M_* 
        - \log M_{\rm *,min}}{\sigma}\right)\right], \label{eq:incomplete}
\end{equation}
where $\rm{erf}$ is the error function and the free parameters are
$f$, $M_{\rm *,min}$ and $\sigma$. This relation allows for the
massive-end stellar mass completeness, $f$, to be less than unity to
take into account the possibility of missing very massive
galaxies. The parameter $M_{\rm *,min}$ sets the stellar mass scale
where on average $f/2$ of the galaxies are selected. This probe using
the SMF measurements from other volume-limited samples provides a
simple and straightforward way to roughly estimate the stellar mass
completeness. However, as the completeness factor is constrained from
small-area deep surveys, one can not take advantage of measuring
accurate stellar mass functions from a very large survey, which are
less impacted by the cosmic variance.

Since the central and satellite galaxies have quite different color
distributions \citep[e.g.,][]{Yang17}, the color selection criteria in
the BOSS observation may thus result in different incompleteness
effects between central and satellite galaxies. Therefore, we adopt
different sets of free parameters in Eq. \ref{eq:incomplete} for
central and satellite galaxies. That is, we have in total six free
parameters to describe the incompleteness functions, three parameters
($f_{\rm I}$, $M_{\rm *,min,I}$ and $\sigma_{\rm I}$) for central
galaxies $c_{\rm I}(M_*)$ and three parameters ($f_{\rm II}$,
$M_{\rm *,min,II}$ and $\sigma_{\rm II}$) for satellite galaxies,
$c_{\rm II}(M_*)$.

The incomplete occupation function of central and satellite galaxies,
i.e., the average number of {\it observed} central and satellite
galaxies with stellar mass $M_{*,1}<M_*<M_{*,2}$ in halos of mass $M$
can be estimated as \citep{Yang12}
\begin{eqnarray}
    \langle N_{\rm c}(M)\rangle&=&\int_{M_{*,1}}^{M_{*,2}} \Phi_{\rm c}(M_*|M) ~c_{\rm I}(M_*) ~dM_* \label{eq:ncen_int}\\
    \langle N_{\rm s}(M)\rangle&=&\int_{M_{*,1}}^{M_{*,2}} \Phi_{\rm s}(M_*|M) ~c_{\rm II}(M_*) ~dM_*,\label{eq:nsat}
\end{eqnarray}
where $\Phi_{\rm s}(M_*|M)$ is the CSMF for the satellite galaxies. 
We assume that satellite galaxies have the same CSMF as the centrals
when they are distinct halos at the last accretion epoch, as
widely used in the subhalo abundance matching algorithm
\citep[see][and references therein]{Yang17}. Then the total satellite
galaxy distribution in halos of mass $M$ can be described as,
\begin{equation}
\Phi_{\rm s}(M_*|M)=\int dM_{\rm acc} \Phi_{\rm c}(M_*|M_{
	\rm acc})n_{\rm s}(M_{\rm acc}|M) \label{eq:csmfs}\,,
\end{equation}
where $M_{\rm acc}$ is the subhalo mass at the last accretion epoch
and $n_{\rm s}(M_{\rm acc}|M)$ is the corresponding subhalo mass
function in host halos of mass $M$. 

Note that within this framework, we have assumed the same SHMR for central and satellite galaxies, which is not exactly true. We have ignored the possible evolution of CSMF between the accretion epoch and the redshift of interest. We also ignore the growth and tidal stripping effects for the satellite galaxy stellar mass after accretion.  As found by recent studies of \cite{Guo16} and \cite{Yang17}, the central and satellite galaxies tend to have somewhat different galaxy-halo relations. As shown in Figure~2 of \cite{Guo16}, the projected 2PCF $\wprp$ would be underestimated when using $M_{\rm acc}$ as the subhalo mass by assuming the same SHMR as the central galaxies. The agreement with the observed $\wprp$ can be improved by adopting different SHMRs for central and satellite galaxies \citep[Figure~6 in][]{Guo16}. 
	
In general, one may assume different SHMRs for central and satellite galaxies, and use the observed $\wprp$ and SMFs to make constraints on both of them. However, assuming different SHMRs for satellite galaxies will introduce another four free parameters ($M_{*,0}$, $M_1$, $\alpha$, and $\beta$). Since the SHMRs of the central and satellite galaxies are not completely independent, the observation would be over-fitted by adopting completely independent SHMRs for central and satellite galaxies. A more consistent and reasonable model is to include the redshift evolution in the central galaxy SHMR and the stellar mass evolution of satellite galaxies after accretion, as those carried out in \cite{Yang12}.  We will defer such a sophisticated model of galaxy stellar mass evolution based on BOSS observation in a subsequent paper. Although we are not modeling the
evolutionary trajectories of satellite galaxies, these effects only
have minimal influence on the total SMF as the majority of galaxies
in these massive galaxy samples are central galaxies \citep{White11,Parejko13,Guo14}.
In addition, as these effects might compromise at certain levels, the
SHMRs obtained from central and satellite galaxies are still quite similar
\citep{Yang17}.

In practice, since we are using halos/subhalos from simulations for
our study, the satellite occupation function in subhalos of mass $M_{\rm acc}$ can
be directly estimated as,
\begin{equation}
  \langle N_{\rm s}(M_{\rm acc})\rangle=\int_{M_{*,1}}^{M_{*,2}}
  \Phi_{\rm c}(M_*|M_{\rm acc})  ~c_{\rm II}(M_*) ~dM_* .\label{eq:nsat_int}
\end{equation}

With these HOD definitions, we can then split the galaxy sample into
different stellar mass ranges, and calculate the clustering
measurements for these subsamples to constrain the ICSMF model
parameters as in \cite{Yang12}. We note that the effect of
$\sigma_{\rm c}$ on the galaxy clustering is automatically taken into
account in Eqs.~\ref{eq:ncen_int} and~\ref{eq:nsat_int}.

\subsubsection{Two-point Correlation Function and Stellar Mass Function}

With the ICSMF for central and satellite galaxies, we measure the
galaxy 2PCF by applying the simulation-based method of \cite{Zheng16}
with the halo catalogs in the BigMDPL simulations, which overcomes the
difficulty of modeling the halo exclusion effect, the scale-dependent
halo bias, and the residual redshift-space distortion effect. By
tabulating the clustering measurements of the halos and subhalos, this
method is equivalent to, but significantly more efficient than,
directly populating halos and subhalos in the simulations with
galaxies using the occupation function, $\langle N_{\rm c}(M)\rangle$
and $\langle N_{\rm s}(M_{\rm acc})\rangle$. In detail, The 3D galaxy
2PCF $\xi({\mathbf r})$ is measured in the BigMDPL simulations as
follows \citep{Zheng16},
\begin{eqnarray}
&&\xi({\mathbf r})=\sum_{i,j}\frac{n_{{\rm h},i}n_{{\rm h},j}}{\bar{n}_{\rm g}^2}
\langle N_{\rm c}(M_i)\rangle\langle N_{\rm c}(M_j)\rangle\xi_{\rm hh}({\mathbf r};M_i,M_j)\nonumber \\
&+&\sum_{i,j}2\frac{n_{{\rm h},i}n_{{\rm s},j}}{\bar{n}_{\rm g}^2}
\langle N_{\rm c}(M_i)\rangle\langle N_{\rm s}(M_{{\rm acc},j}) \rangle\xi_{\rm hs}({\mathbf r};M_i,M_{{\rm acc},j})\nonumber \\
&+&\sum_{i,j}\frac{n_{{\rm s},i}n_{{\rm s},j}}{\bar{n}_{\rm g}^2}
\langle N_{\rm s}(M_{{\rm acc},i})\rangle\langle N_{\rm s}(M_{{\rm
  acc},j})\rangle \xi_{\rm ss}({\mathbf r};M_{{\rm acc},i},M_{{\rm acc},j})\nonumber\\ \label{eq:xicc_sim}
\end{eqnarray}
where $n_{\rm h}(M)$ and $n_{\rm s}(M_{\rm acc})$ are the halo mass
function and subhalo mass function at the last accretion epoch,
respectively. The predicted galaxy number density $\bar{n}_{\rm g}$
can be calculated as,
\begin{equation}
\bar{n}_{\rm g}=\sum_i\left[\langle N_{\rm c}(M_i)\rangle n_{\rm
    h}(M_i)+ \langle N_{\rm s}(M_{{\rm acc},i})\rangle n_{\rm s}(M_{{\rm acc},i})\right]. \label{eq:nc}
\end{equation}  
The 3D 2PCFs $\xi_{\rm hh}({\mathbf r};M_i,M_j)$,
$\xi_{\rm hs}({\mathbf r};M_i,M_{{\rm acc},j})$, and
$\xi_{\rm ss}({\mathbf r};M_{{\rm acc},i},M_{{\rm acc},j})$ are the
tabulated 2PCFs of the halo-halo, halo-subhalo, and subhalo-subhalo
pairs between the different mass bins from the simulation. With the help of halo and
subhalo catalogs, we can accurately measure $\xi({\mathbf r};M_i,M_j)$
directly from the simulations, rather than applying any theoretical
models for the halo bias $b_{\rm h}(M)$ \citep{Mo96}.

To reduce the effect of redshift-space distortion (RSD) caused by
galaxy peculiar velocities, we focus on the measurements of the
projected 2PCF $w_{\rm p}(r_{\rm p})$ \citep{Davis83}, defined as,
\begin{equation}
w_{\rm p}(r_{\rm p})=2\int_{0}^{r_{\rm \pi,max}} \xi(r_{\rm p},r_{\rm\pi})dr_{\rm\pi}, \label{eq:wp}
\end{equation}
where $r_{\rm\pi}$ and $r_{\rm p}$ are the separations of galaxy pairs
along and perpendicular to the line-of-sight (LOS). $r_{\rm \pi,max}$
is the maximum LOS distance to achieve the best signal-to-noise ratio.

The observed (incomplete) galaxy SMF of the BOSS sample can
be predicted as,
\begin{eqnarray}
\Phi(M_*)&=&\int dM \Phi_{\rm c}(M_*|M) c_{\rm I}(M_*) n_{\rm h}(M)\nonumber\\
&+&\int dM_{\rm acc} \Phi_{\rm c}(M_*|M_{\rm acc}) c_{\rm II}(M_*) n_{\rm s}(M_{\rm acc})
\end{eqnarray}

In summary, we have four free parameters ($M_{*,0}$, $M_1$, $\alpha$,
and $\beta$) for the stellar-halo mass relation (Eq.~\ref{eq:smhm})
and another six parameters ($f_{\rm I}$, $M_{\rm *,min,I}$,
$\sigma_{\rm I}$, $f_{\rm II}$, $M_{\rm *,min,II}$ and
$\sigma_{\rm II}$) for the incompleteness component
(Eq.~\ref{eq:incomplete}). The predictions of $\Phi(M_*)$ and
$w_{\rm p}(r_{\rm p})$ can then be compared with those measured in the
observed galaxy sample in order to constrain the best-fitting model
parameters.

With the best-fitting model parameters, we can infer the intrinsic
galaxy total SMF $\tilde{\Phi}(M_*)$ at each redshift
interval as,
\begin{eqnarray}
\tilde{\Phi}(M_*)&=&\int dM \Phi_{\rm c}(M_*|M)  n_{\rm h}(M)\nonumber\\
&+&\int dM_{\rm acc} \Phi_{\rm c}(M_*|M_{\rm acc}) n_{\rm s}(M_{\rm acc})\,.
\end{eqnarray}

\begin{figure*}
\centering
\includegraphics[width=0.8\textwidth]{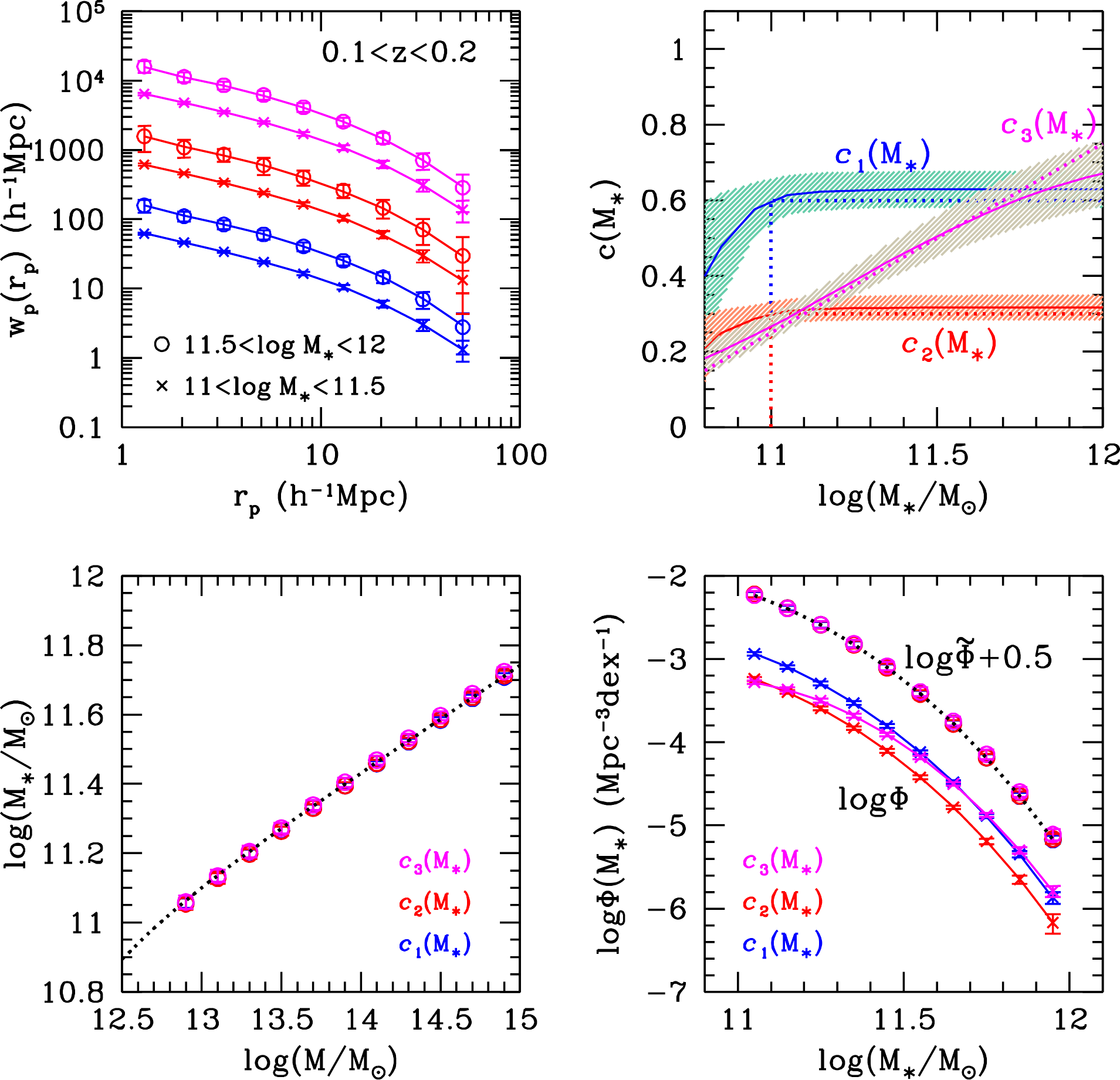}
\caption{Tests of our method using mock galaxy catalogs. We
  constructed three mock catalogs with different selection functions
  of Eqs.~\ref{eq:fh1}, \ref{eq:fh2} and \ref{eq:fh3}, displayed as
  the blue, red, and magenta symbols (and lines), respectively. Top
  Left: measurements of $w_{\rm p}(r_{\rm p})$ with the circles and
  crosses for the stellar mass ranges of $10^{11}$--$10^{11.5}\msun$
  and $10^{11.5}$--$10^{12}\msun$, respectively. The measurements for
  the $c_2(M_*)$ and $c_3(M_*)$ models are shifted upward by
  1~dex and 2~dex for clarity, respectively. The best-fitting models
  are shown as the solid lines. Top Right: the three different
  selection functions, with the dotted lines as the input models and
  the solid lines with shaded regions as the median and 1$\sigma$ uncertainties of the best-fitting models. Bottom Left: comparisons of the
  stellar-halo mass relations, with the circles for the best-fitting
  models and the dotted line for the input model. Bottom Right: the
  observed and total galaxy SMFs, with the crosses for the observed
  galaxy SMFs $\Phi(M_*)$ and the circles for the predicted total SMFs
  $\tilde{\Phi}(M_*)$ from the best-fitting models. The solid lines
  are the best-fitting models for $\Phi(M_*)$. The black dotted line
  is for the input model of $\tilde{\Phi}(M_*)$. The measurements for
  $\tilde{\Phi}(M_*)$ are shifted upward by 0.5~dex for clarity.  }
	\label{fig:mock}
\end{figure*}

\begin{figure*}
\centering
\includegraphics[width=0.8\textwidth]{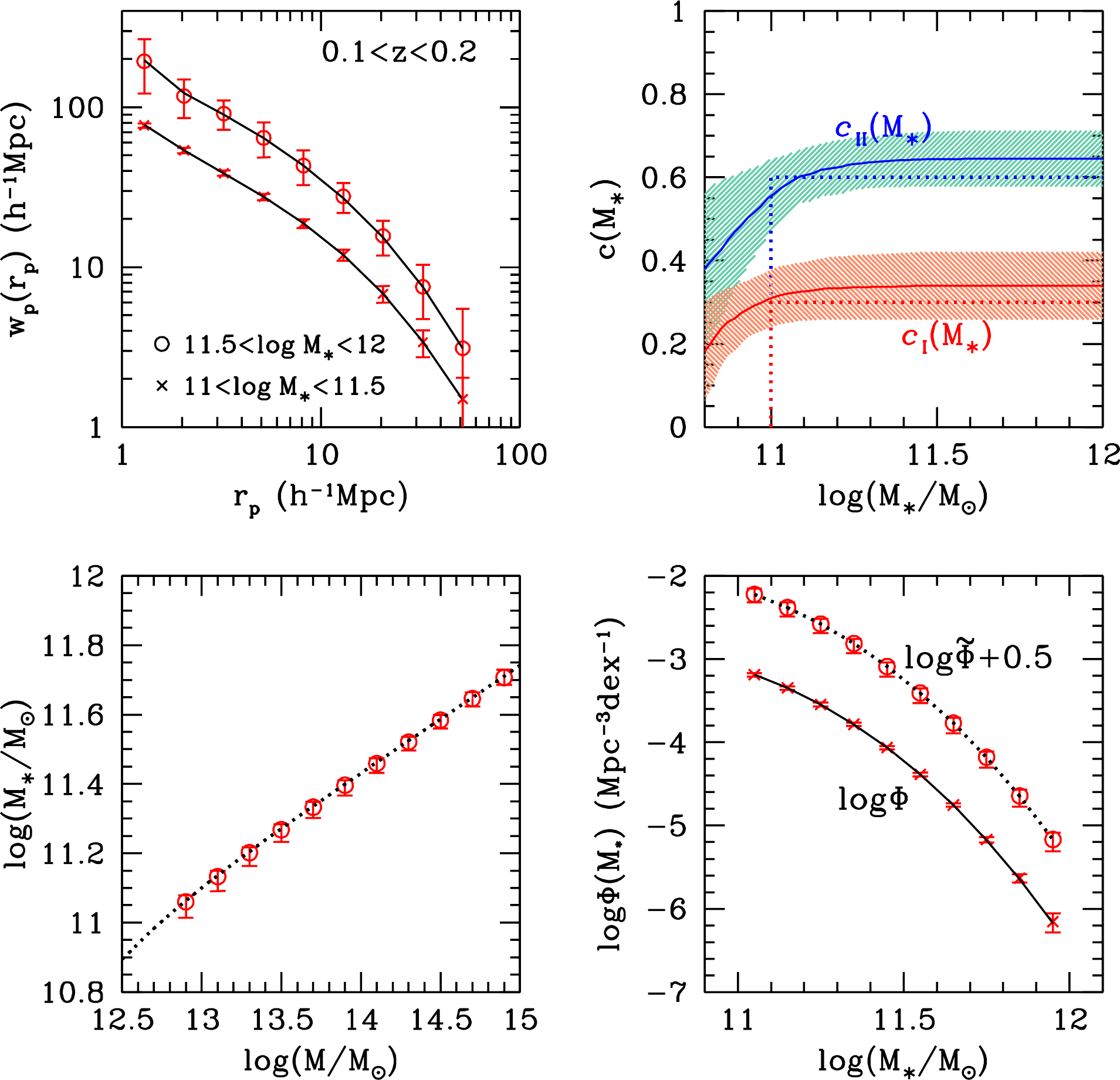}
\caption{Similar to Figure~\ref{fig:mock}, but for the case of
  different completeness functions for central and satellite galaxies,
      i.e., $c_{\rm I}(M_*)=0.3$ and $c_{\rm II}(M_*)=0.6$.}
	\label{fig:mock_compcs}
\end{figure*}

\subsection{Observational Measurements}
\label{sec:obs_meas}

With the above listed models for the galaxy projected 2PCFs and SMFs,
we can use the related observational measurements to constrain the
model parameters.

We measure the projected 2PCF $\wprp$ for BOSS galaxies through
the Landy--Szalay estimator \citep{Landy93}. In practice, we integrate
$\xi(r_{\rm p},r_{\rm\pi})$ in Eq.~(\ref{eq:wp}) to a maximum LOS
distance of $r_{\pi,{\rm max}}=100\mpchi$ to achieve the best
signal-to-noise ratio, and the same value is adopted in the model
calculation. We choose logarithmic $r_{\rm p}$ bins with a width
$\Delta\log r_{\rm p}=0.2$ from $1$ to $63.1\mpchi$, and linear
$r_{\rm\pi}$ bins of width $\Delta r_{\rm\pi}=2\mpchi$ from 0 to
100$\mpchi$. We measure the projected 2PCFs for the two stellar mass
ranges of $10^{11}\msun<M_*<10^{11.5}\msun$ and
$10^{11.5}\msun<M_*<10^{12}\msun$.

The observed galaxy SMF is measured for galaxy stellar mass in the
range of $10^{11}$ to $10^{12}\msun$ with a logarithmic width of
$\Delta\log M_*=0.1$. Therefore, we have 28 data points in total, with
18 for $\wprp$ and 10 for $\Phi(M_*)$. We estimate the error
covariance matrices for $w_{\rm p}(r_{\rm p})$ and $\Phi(M_*)$ using
the jackknife re-sampling technique with 100 subsamples
\citep{Guo13,Guo14}. We note that the cross-covariance between the
$w_{\rm p}(r_{\rm p})$ measurements for the two stellar mass bins are
also taken into account in the covariance matrix. The jackknife
re-sampling method provides a reasonable way to estimate the sample
variance effect \citep[see agreement with the errors estimated from
mock catalogs in the Appendix B of][]{Guo13}.

\subsection{Model Constraints}
\label{sec:mod_cons}

In order to fully explore the probability distribution of the model
parameters, we apply a Markov Chain Monte Carlo (MCMC) method.  The
likelihood surface is determined by $\chi^2$, contributed by the
observed galaxy stellar mass function $\Phi(M_*)$ and the projected
2PCF $\wprp$,
\begin{eqnarray}
\chi^2&=&\chi^2_{\wrp}+\frac{(\Phi-\Phi^*)^2}{\sigma_{\Phi}^2}\\
\chi^2_{\wrp}&=& (\mathbf{\wrp}-\mathbf{\wrp^*})^T
                 \mathbf{C}_{\wrp}^{-1} (\mathbf{\wrp}-\mathbf{\wrp^*}),
\end{eqnarray}
where $\mathbf{C}_{\wrp}$ is the full error covariance matrix of
$\wprp$. The quantity with (without) a superscript `$*$' is the one
from the data (model). The degree of freedom (dof) of the model is 18,
i.e., ${\rm dof}=28-10$.

Here we only adopt the diagonal elements of the covariance matrix of
$\Phi$, because the uncertainties from systematic effects of stellar
mass measurements are hard to estimate \citep{Mitchell13}. The contribution of Poisson
noise to the observed SMF is added in quadrature to $\sigma_{\Phi}$.

\section{Test the performance of our method}\label{sec:test}

\begin{figure*}
 \centering
 \includegraphics[width=0.8\textwidth]{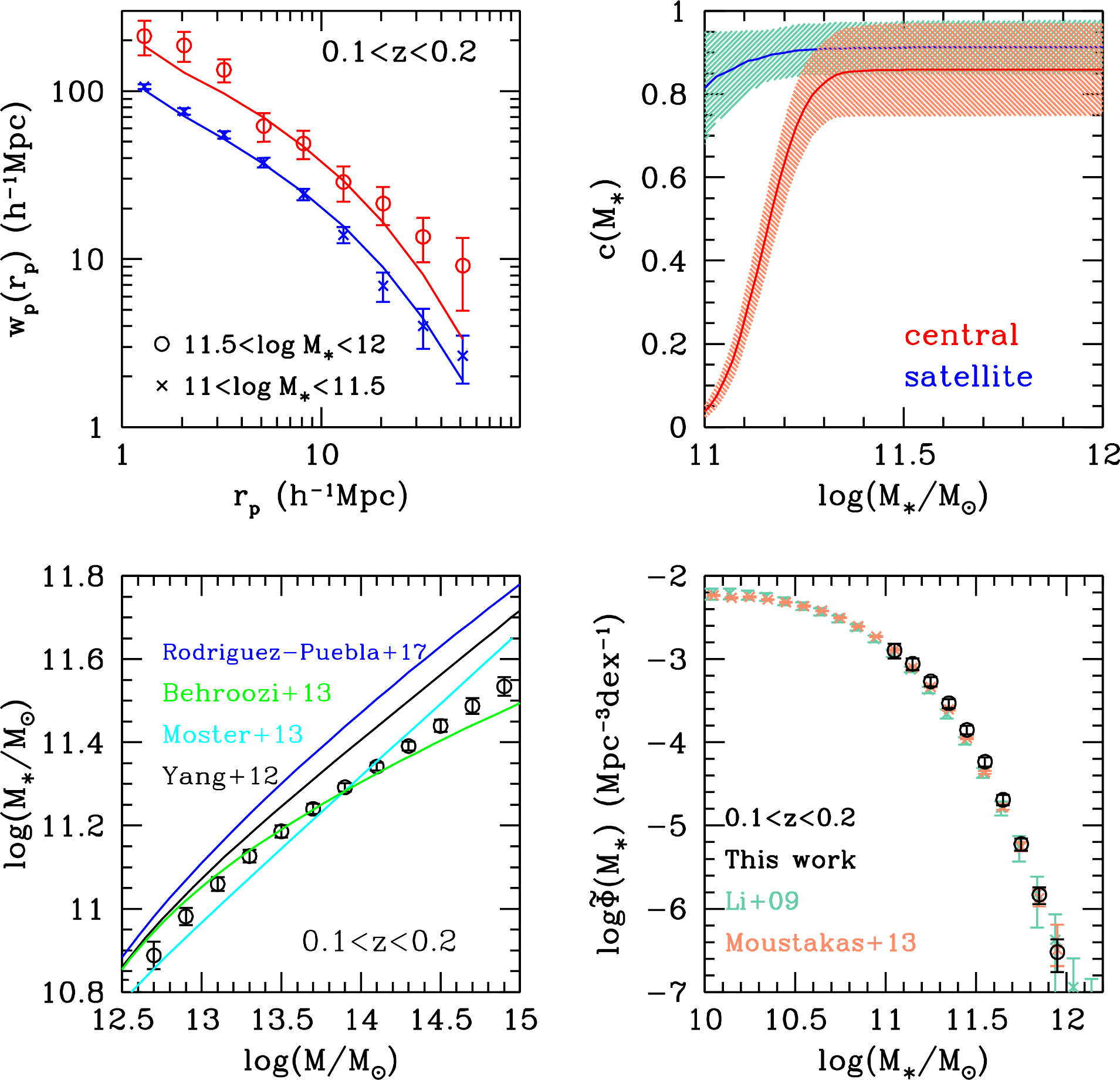}
 \caption{Similar to Figure~\ref{fig:mock}, but for the BOSS galaxies
   at $0.1<z<0.2$. We show the SHMR of our best-fitting model as the
   circles in the bottom left panel. In comparisons, those of
   \cite{Yang12}, \cite{Moster13}, \cite{Behroozi13} and
   \cite{Rodriguez-Puebla17} are shown as lines of different colors.
   In the bottom right panel, we compare our measurement of the galaxy
   total SMF (circles) with those obtained by \cite{Li09} and
   \cite{Moustakas13} (crosses with different colors) }
 \label{fig:sdss}
\end{figure*}

\subsection{Tests with mock catalogs}
\label{sec:mock_test}

Before we apply our method to the BOSS galaxies, we perform a validity
test on a mock galaxy catalog in the redshift range of $0.1<z<0.2$
that has the same geometry as the BOSS sample. We first assign each
halo in the BigMDPL simulation a value of galaxy stellar mass
according to the following model parameters of Eq.~\ref{eq:smhm},
\begin{equation}
\log M_{*,0}=10.459, \log M_1=10.844, \alpha= 0.309, \beta=8.077, \nonumber
\end{equation}
which are adopted from \cite{Yang12} as the best-fitting parameters at
the simulation output redshift of $z=0.152$ \citep[see][for more
parameters for different sets of cosmologies]{Yang12}. We note that
satellite galaxies are included in the mock catalogs by applying the
SHMR with the subhalo mass at the accretion epoch. Once the dark
matter halos and subhalos are populated with galaxies of different
stellar masses, we then place a virtual observer at the center of the
simulation volume and calculate the right ascension, declination and
redshift for each galaxy. The galaxy peculiar velocity is taken into
account when calculating the redshift.

We first test our method for the case of the same stellar mass
completeness function $c(M_*)$ for central and satellite galaxies,
i.e., $c_{\rm I}(M_*)=c_{\rm II}(M_*)$. We constructed three mock
catalogs by applying the following three simple selection models on
the central and satellite galaxies,
\begin{eqnarray}
c_1(M_*)&=&0.6,   \label{eq:fh1}\\
c_2(M_*)&=&0.3,  \label{eq:fh2}\\
c_3(M_*)&=&(\log M_*-10.5)/2, \label{eq:fh3}
\end{eqnarray}
The models $c_1(M_*)$ and $c_2(M_*)$ are random selections with
different rates. The model $c_3(M_*)$ is a linear selection with
higher rates for more massive galaxies.
\begin{figure*}
	\centering
	\includegraphics[width=0.8\textwidth]{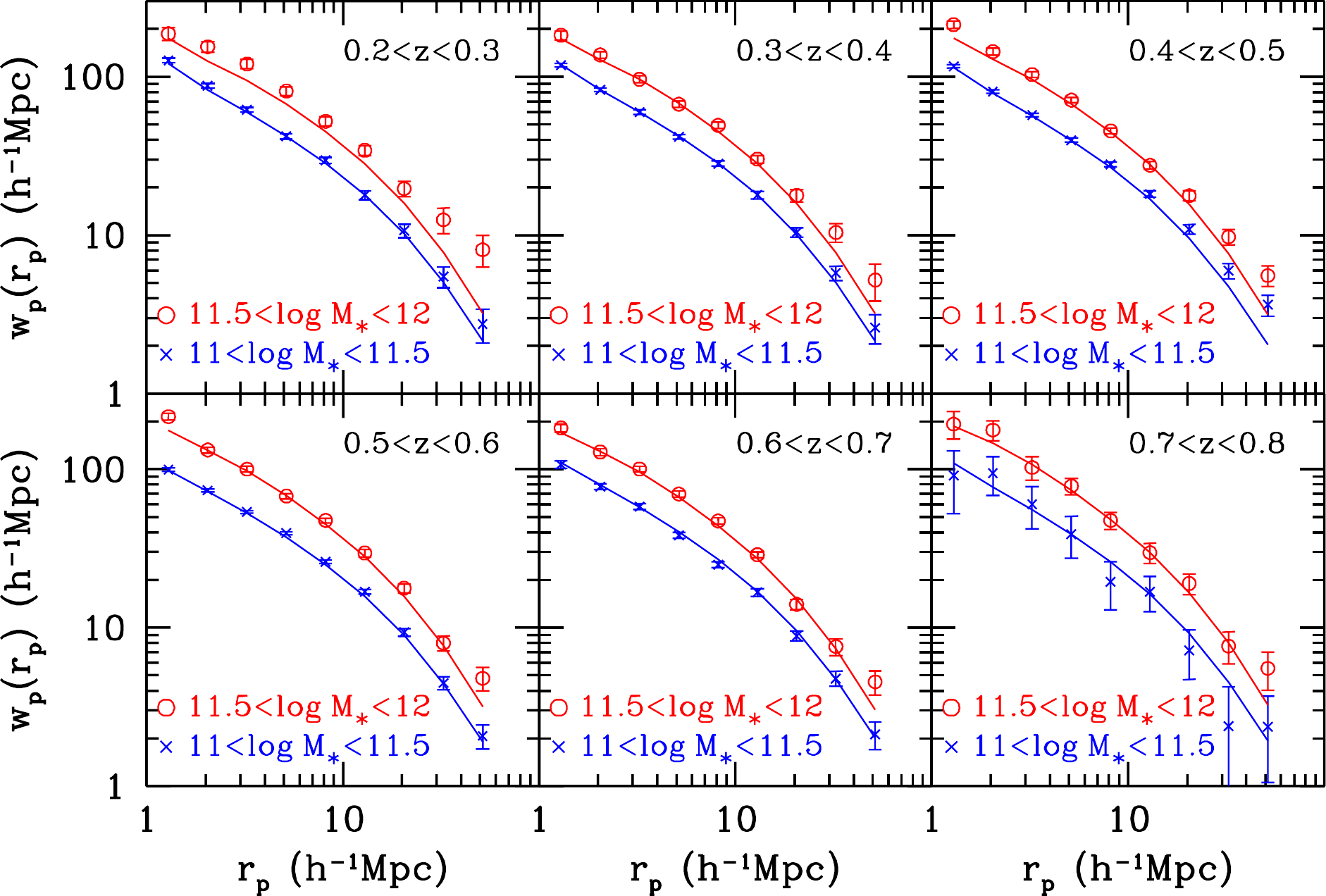}
	\caption{Projected 2PCFs for the BOSS galaxies at redshifts from 0.2
		to 0.8. The blue and red symbols are for the stellar mass ranges of
		$10^{11}$--$10^{11.5}\msun$ and $10^{11.5}$--$10^{12}\msun$,
		respectively. The solid lines are our best-fitting models.  }
	\label{fig:wp}
\end{figure*}

We then measure $w_{\rm p}(r_{\rm p})$ and $\Phi(M_*)$ for these mock
catalogs and run the MCMC chains to find the best-fitting model
parameters. The results are displayed in Figure~\ref{fig:mock}, with
the blue, red and magenta symbols (and lines) for the three mock
catalogs with $c_1(M_*)$, $c_2(M_*)$ and $c_3(M_*)$, respectively. The
top left panel shows the projected 2PCFs $w_{\rm p}(r_{\rm p})$, with
the crosses and circles for the stellar mass ranges of
$10^{11}$--$10^{11.5}\msun$ and $10^{11.5}$--$10^{12}\msun$,
respectively. For clarity, the measurements for the $c_2(M_*)$ and $c_3(M_*)$
models are shifted upward by 1~dex and 2~dex,
respectively. The completeness functions are shown in the top right
panel, where the dotted lines are for the input models and the solid
lines with shaded regions are the average of the best-fitting models with
1$\sigma$ uncertainties. As can be seen, the fiducial $c(M_*)$ are
very nicely recovered for all the three different selection models. As we only use galaxies with stellar
masses larger than $10^{11}\msun$ for both $w_{\rm p}(r_{\rm p})$ and $\Phi(M_*)$, the discrepancies shown for $M_*<10^{11}\msun$ do not have any effect on the results.

The stellar-halo mass relations from the best-fitting models are shown
as the circles in the bottom left panel, where the dotted line is the
input model. The input model parameters are also very well recovered
from the best-fitting models. The observed galaxy SMFs are
shown as the crosses with different colors in the bottom right panel,
where the best-fitting models are shown as the solid lines. The
predicted galaxy total SMFs from the best-fitting models are displayed
as the circles, which are shifted upward by 0.5~dex for clarity. The
fiducial model (black dotted line) is successfully recovered from our
best-fitting models.

Before we apply our ICSMF model to the BOSS observation, it is
important to check to what extent the true incompletenesses for
central and satellite galaxies can be recovered if they are
different. To this end, we input $c_{\rm I}(M_*)=0.3$ for central
galaxies, $c_{\rm II}(M_*)=0.6$ for satellite galaxies, and perform
the same procedure above. As shown in Figure~\ref{fig:mock_compcs}, both the input completeness
functions, the SHMR, as well as the total galaxy SMF are very well
recovered, demonstrating the validity of our ICSMF model to accurately
predict the galaxy SMFs from incomplete samples with complicated
target selections. In the following sections, we will use separate
completeness functions for central and satellite galaxies to model the
real BOSS galaxy samples.

\begin{figure*}
	\centering
	\includegraphics[width=0.8\textwidth]{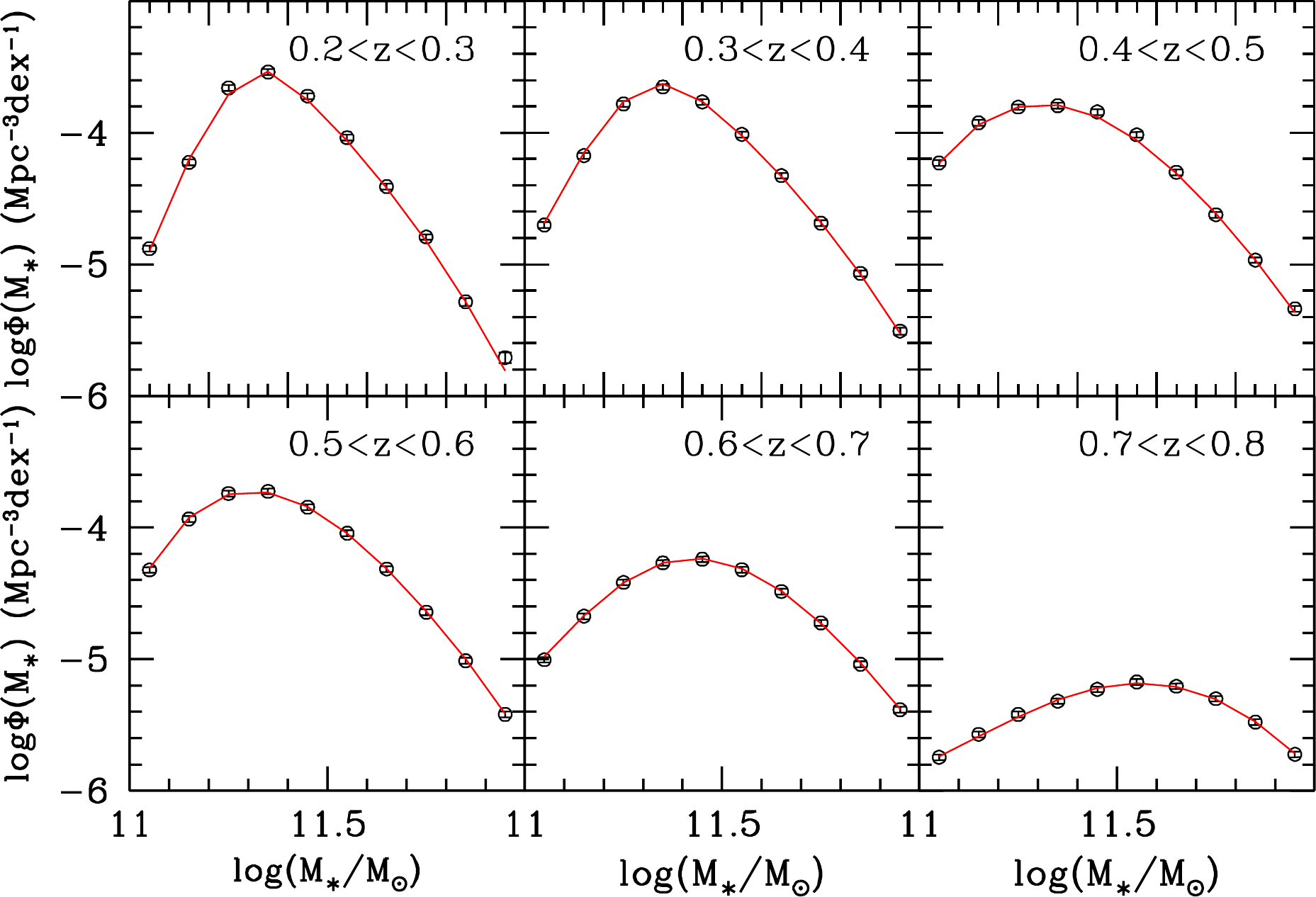}
	\caption{Observed galaxy SMF (circles with errors),
		$\Phi(M_*)$, in units of $\rm{Mpc}^{-3}\rm{dex}^{-1}$, in
		different redshift ranges from $z=0.2$ to $0.8$. The solid
		lines are our best-fitting models.  }
	\label{fig:smf}
\end{figure*}
\subsection{Test with BOSS LOWZ galaxy sample}

As the BOSS LOWZ galaxy sample also covers the redshift range of
$0.1<z<0.2$, where accurate galaxy SMFs have been measured in
literature using the SDSS DR7 Main galaxy sample. As a consistency
check, we also apply our method to the BOSS galaxies in this redshift
range and compare with literature. 

We show in Figure~\ref{fig:sdss} the modeling results for the $\wprp$
(top left panel), $c(M_*)$ (top right panel), SHMR (bottom left
panel), and the total galaxy SMF (bottom right panel) at $0.1<z<0.2$
as in Figure~\ref{fig:mock}. The low-redshift sample of BOSS is more
than 85\% complete for $M_*>10^{11.4}\msun$. 

In the bottom left panel, we compare the SHMR for our best-fitting
model (circles) with the predictions of different theoretical models
of \cite{Yang12}, \cite{Behroozi13}, \cite{Moster13}, and
\cite{Rodriguez-Puebla17} (lines of different colors). The high-mass
end slope of the SHMR is not well constrained in previous literature,
where significant differences exist among the models. These
differences mainly come from three origins: (1) the stellar mass
functions in these studies are somewhat different; (2) the
cosmological parameters adopted are different; (3) the scatter
$\sigma_c$ used in these studies are somewhat different as well.  Our
measurements here agree best with the model of \cite{Behroozi13} at
this redshift interval.
 
Our predicted galaxy total SMF from the BOSS sample is shown as the
black circles in the bottom right panel. We also compare them with the
galaxy SMFs measured in \cite{Li09} using SDSS DR7 and
\cite{Moustakas13} using the joint sample of SDSS and Galaxy Evolution
Explorer \citep[GALEX;][]{Martin05}. We use the stellar mass
measurements estimated with the $r$-band model magnitude from
\cite{Li09} to account for the fact that the typical r-band Petrosian
magnitudes would result in the underestimation of the galaxy stellar
mass \citep{Guo10}. We add a constant offset of $-0.05$~dex to the
stellar mass of the \cite{Moustakas13} to account for the Flexible
Stellar Population Synthesis (FSPS) models used in their stellar mass
estimates (see their Figure~19 for comparisons of different SPS
models). We find very good agreement between our measurements and
those from the literature.
 
The above two sets of tests demonstrate that our method is robust in
recovering the missing fraction of galaxies, as well as providing
unbiased measurements of the galaxy SMFs and SHMRs from large-scale surveys
with complicated target selections. In the following sections, we will
apply our method to the BOSS galaxies in $0.2<z<0.8$ to predict the
galaxy total SMFs and SHMRs.

\section{Results}\label{sec:results}

\subsection{ICSMF model constraints}

We measure the projected 2PCFs and the incomplete SMFs from BOSS
observation within redshift range of $0.2<z<0.8$ according to the methods
outlined in section \ref{sec:obs_meas}. Here galaxies are divided into
different samples according to the criteria listed in Table
\ref{tab:data}.  These measurements are then used to constrain the
ICSMF model parameters using the algorithm outlined in section
\ref{sec:mod_cons}.

We show in Figure~\ref{fig:wp} the projected 2PCF measurements for the
BOSS galaxies at $0.2<z<0.8$. The blue and red symbols are for
galaxies in the stellar mass ranges of $10^{11}$--$10^{11.5}\msun$ and
$10^{11.5}$--$10^{12}\msun$, respectively. Our best-fitting models are
shown as the solid lines of different colors. The clustering amplitude
of $\wprp$ for the more massive sample is about twice that of the
lower mass sample, which is consistent with the galaxy bias
measurements of \cite{Tinker17}. As these BOSS galaxies generally live
in halos of masses larger than $10^{12}\msun$, the halo bias is
quickly increasing with mass for these massive halos
\citep{Mo96,Tinker05}. Our best-fitting models reasonably fit all the
observed $\wprp$, except for the high mass sample at
$0.2<z<0.3$. While the best-fitting $\chi^2$ (41.7) for this sample is
still reasonable for a degree-of-freedom of 18, the slight deviation
is possibly related to the assumption of the same SHMR for central and
satellite galaxies and the constant scatter $\sigma_{\rm c}$.  The
current BOSS measurements are not accurate enough to fully constrain
those sophisticated models with more freedom.

Figure~\ref{fig:smf} shows our best-fitting models for the observed
galaxy SMFs at different redshifts, with the circles for the measured
galaxy SMFs in BOSS and solid lines for our best-fitting models. The agreement between the data and models for galaxy samples
at different redshifts are remarkably good, implying that the sample
selection functional froms used in this study are suitable.

\begin{figure*}
	\centering
	\includegraphics[width=0.8\textwidth]{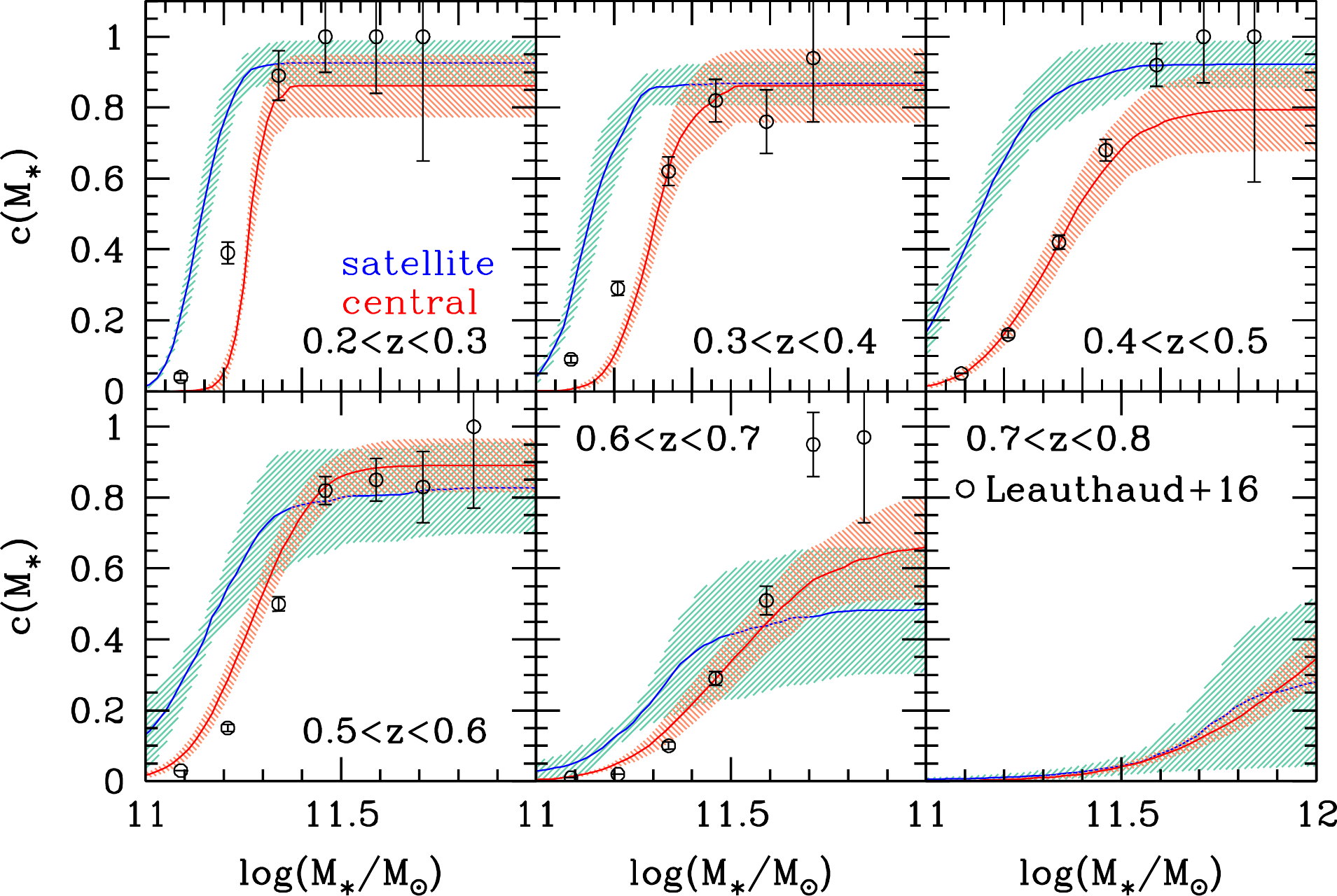}
	\caption{Stellar mass completeness functions for galaxies at different
		redshift ranges. The solid lines with the shaded area are the median and $1\sigma$ uncertainties of the best-fitting models. The red and blue lines are for the central and satellite galaxies, respectively. The cicles with errors are the estimates of the stellar mass completeness of all galaxies at $0.2<z<0.7$ from \cite{Leauthaud16}.}
	\label{fig:ncen}
\end{figure*}

The best-fitting stellar mass completeness functions for the different
galaxy samples are shown in Figure~\ref{fig:ncen}, with the red and
blue solid lines for central and satellite galaxies, respectively. The
shaded area represent the model uncertainties. The BOSS galaxy samples
are more than 80\% complete for massive galaxies of $M>10^{11.5}\msun$
at $z<0.6$, but the completeness decreases very fast for lower mass
galaxies, causing the galaxy SMF to decrease at the low mass end in
Figure~\ref{fig:smf}. At higher redshifts of $z>0.7$, even those very
massive galaxies are only about 30\% complete. For comparison, we also display the completeness estimates from \cite{Leauthaud16} as the circles in the redshift range of $0.2<z<0.7$. As their estimates are for the whole BOSS sample, they are more comparable to the completeness functions of our central galaxies, because the central galaxies dominate the BOSS sample. Our predictions are generally in good agreement with their estimates at all redshifts. 

The completeness of satellite galaxies at the massive end,
$f_{\rm II}$, is not as well constrained as that of central galaxies, as
seen from the 1$\sigma$ distribution. It is related to the fact that
central galaxies dominate the massive end of the SMF
\citep{Yang09}. But at the low mass end, the completeness of satellite
galaxies is quite well constrained and generally larger than than that
of the central galaxies. It is caused by the $i$-band sliding cut
($i<19.86+1.6(d_{\perp}-0.8)$) of the BOSS CMASS target selections to
remove the fainter and bluer galaxies \citep[see e.g., Figure~1
of][]{Guo13}, while the low-mass red galaxies have higher satellite
fractions \citep[see e.g.,][]{Zehavi11,Guo14,Saito16,Yang17}. Therefore, the
color selection of the BOSS galaxies samples at a given stellar mass
is also taken into account in our separate modeling of the
completeness functions for the central and satellite galaxies.

\begin{table*}
	\caption{Best-fitting Model Parameters} \label{tab:model}
	\centering
	\begin{tabular}{lrrrrrrr}
		\hline
		Parameter & $0.1<z<0.2$ & $0.2<z<0.3$ & $0.3<z<0.4$ & $0.4<z<0.5$ & $0.5<z<0.6$ & $0.6<z<0.7$ & $0.7<z<0.8$ \\
		\hline
		$\chi^2/{\rm dof}$ & $19.96/18$ & $41.76/18$ & $13.92/18$ & $44.01/18$ & $34.91/18$ & $21.75/18$ & $10.20/18$ \\
		\hline
		
		$\log M_{*,0}$ & $10.74^{+0.06}_{-0.20}$ & $10.45^{+0.12}_{-0.01}$ & $10.38^{+0.21}_{-0.22}$ & $10.36^{+0.06}_{-0.17}$ & $10.46^{+0.08}_{-0.23}$ & $10.37^{+0.19}_{-0.22}$ & $9.92^{+0.51}_{-0.06}$ \\
		
		$\log M_1$ & $11.44^{+0.01}_{-0.42}$ & $11.04^{+0.25}_{-0.01}$ & $11.12^{+0.35}_{-0.39}$ & $11.18^{+0.10}_{-0.28}$ & $11.35^{+0.11}_{-0.33}$ & $11.34^{+0.08}_{-0.27}$ & $10.46^{+0.93}_{-0.08}$ \\
		
		$\alpha$ & $0.23^{+0.03}_{-0.02}$ & $0.33^{+0.02}_{-0.01}$ & $0.38^{+0.01}_{-0.03}$ & $0.42^{+0.02}_{-0.01}$ & $0.41^{+0.03}_{-0.01}$ & $0.47^{+0.02}_{-0.09}$ & $0.48^{+0.05}_{-0.06}$ \\
		
		$\beta$ & $7.89^{+0.46}_{-0.05}$ & $8.00^{+0.06}_{-0.29}$ & $8.25^{+0.24}_{-0.81}$ & $8.49^{+0.06}_{-0.75}$ & $7.49^{+0.89}_{-0.19}$ & $7.30^{+1.18}_{-0.28}$ & $8.13^{+0.41}_{-0.62}$ \\
		
		$f_{\rm I}$ & $0.94^{+0.03}_{-0.20}$ & $0.93^{+0.02}_{-0.16}$ & $0.93^{+0.04}_{-0.17}$ & $0.82^{+0.09}_{-0.14}$ & $0.87^{+0.09}_{-0.05}$ & $0.51^{+0.37}_{-0.01}$ & $0.53^{+0.38}_{-0.01}$ \\
		
		$\log M_{\rm *,min,I}$ & $11.13^{+0.03}_{-0.01}$ & $11.28^{+0.01}_{-0.03}$ & $11.29^{+0.02}_{-0.01}$ & $11.33^{+0.02}_{-0.02}$ & $11.28^{+0.01}_{-0.04}$ & $11.43^{+0.18}_{-0.02}$ & $11.90^{+0.18}_{-0.01}$ \\
		
		$\sigma_{\rm I}$ & $0.11^{+0.02}_{-0.01}$ & $0.07^{+0.01}_{-0.04}$ & $0.11^{+0.02}_{-0.01}$ & $0.23^{+0.01}_{-0.04}$ & $0.21^{+0.01}_{-0.06}$ & $0.23^{+0.13}_{-0.01}$ & $0.39^{+0.11}_{-0.01}$\\
		
		$f_{\rm II}$ & $0.94^{+0.03}_{-0.10}$ & $0.99^{+0.01}_{-0.14}$ & $0.96^{+0.01}_{-0.15}$ & $0.97^{+0.02}_{-0.11}$ & $0.96^{+0.01}_{-0.26}$ & $0.55^{+0.11}_{-0.22}$ & $0.23^{+0.36}_{-0.17}$ \\
		
		$\log M_{\rm *,min,II}$ & $10.56^{+0.35}_{-0.09}$ & $11.15^{+0.02}_{-0.03}$ & $11.14^{+0.03}_{-0.04}$ & $11.13^{+0.02}_{-0.04}$ & $11.18^{+0.01}_{-0.08}$ & $11.33^{+0.07}_{-0.07}$ & $11.92^{+0.14}_{-0.47}$ \\
		
		$\sigma_{\rm II}$ & $0.30^{+0.12}_{-0.25}$ & $0.09^{+0.02}_{-0.02}$ & $0.12^{+0.02}_{-0.05}$ & $0.17^{+0.09}_{-0.03}$ & $0.13^{+0.28}_{-0.01}$ & $0.36^{+0.06}_{-0.23}$ & $0.68^{+0.17}_{-0.47}$\\								
		\hline
	\end{tabular}
	
	\medskip
	All masses are in units of $\msun$.
\end{table*}

\subsection{Central Galaxy Stellar-Halo Mass Relation}
\begin{figure*}
	\centering
	\includegraphics[width=0.8\textwidth]{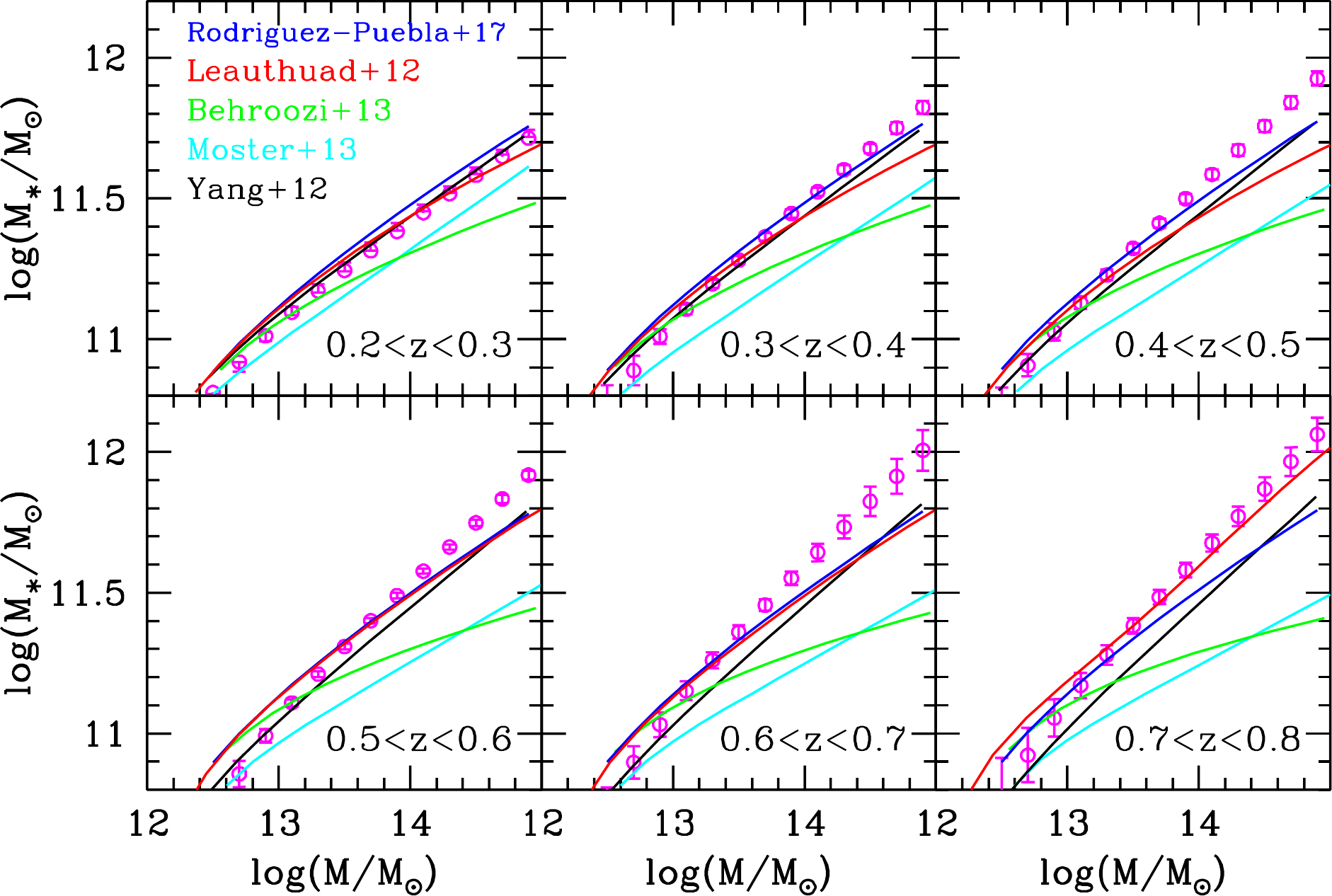}
	\caption{Comparisons of our best-fitting models of central galaxy
		SHMRs (circles with errors) to those of \cite{Leauthaud12},
		\cite{Yang12}, \cite{Behroozi13}, \cite{Moster13}, and
		\cite{Rodriguez-Puebla17} (lines of different colors).  }
	\label{fig:smhm}
\end{figure*}

We show in Figure~\ref{fig:smhm} the best-fitting central galaxy SHMRs
for the different redshift samples (circles with errors), in
comparison to five other different theoretical models of
\cite{Leauthaud12}, \cite{Yang12}, \cite{Behroozi13}, \cite{Moster13},
and \cite{Rodriguez-Puebla17} (lines of different colors). For fair
comparisons, we have corrected the predicted halo masses for the
different cosmologies used in these models. The halos in
\cite{Leauthaud12}, \cite{Yang12} and \cite{Moster13} are defined as
200 times the background mass density, 180 times the background
density and 200 times the critical density, respectively. We correct
these definitions to our definition of the virial halo mass in the
\texttt{ROCKSTAR} halo finder using the average offsets in the BigMDPL
simulation. At $z>0.5$, the halo definition of 200 times the
background density is very similar to the virial mass definition, with
a correction at the level of about 0.01~dex.

Generally, we find a steeper slope of the central galaxy SHMR for
these massive galaxies at $z>0.3$, compared to other models. In these
previous models, the galaxy SHMR is generally derived by fitting
models to a set of different observations of the galaxy SMFs,
clustering measurements, or galaxy weak lensing measurements at
different redshifts. Because of the different systematics and
statistical errors in the estimated galaxy stellar masses, fitting one
SHMR model to different galaxy survey data would possibly generate
inconsistent estimates of the corresponding halo masses at different
redshifts. Therefore, we expect the existence of reasonable systematic
offsets, although we have largely corrected for the assumptions about
the IMF, SPS model and the dust attenuation model following the
various offset values suggested in \cite{Rodriguez-Puebla17}.

As the high-mass end galaxy SMF measurements would be significantly
affected by the Malmquist bias, the shape of the galaxy SHMR may
somewhat dependent on the assumption of $\sigma_{\rm c}$. We will
discuss the effect of different $\sigma_{\rm c}$ values in
Section~\ref{sec:discussion}. We show in Table~\ref{tab:model} the
best-fitting model parameters. We note that the low-mass end slope
$\alpha+\beta$ is not well constrained by our samples as we are only
using galaxies more massive than $10^{11}\msun$. The high-mass end
slope $\alpha$ varies from about $1/4$ to $1/2$ from $z=0.1$ to $z=0.8$.

\begin{figure*}
	\centering
	\includegraphics[width=0.8\textwidth]{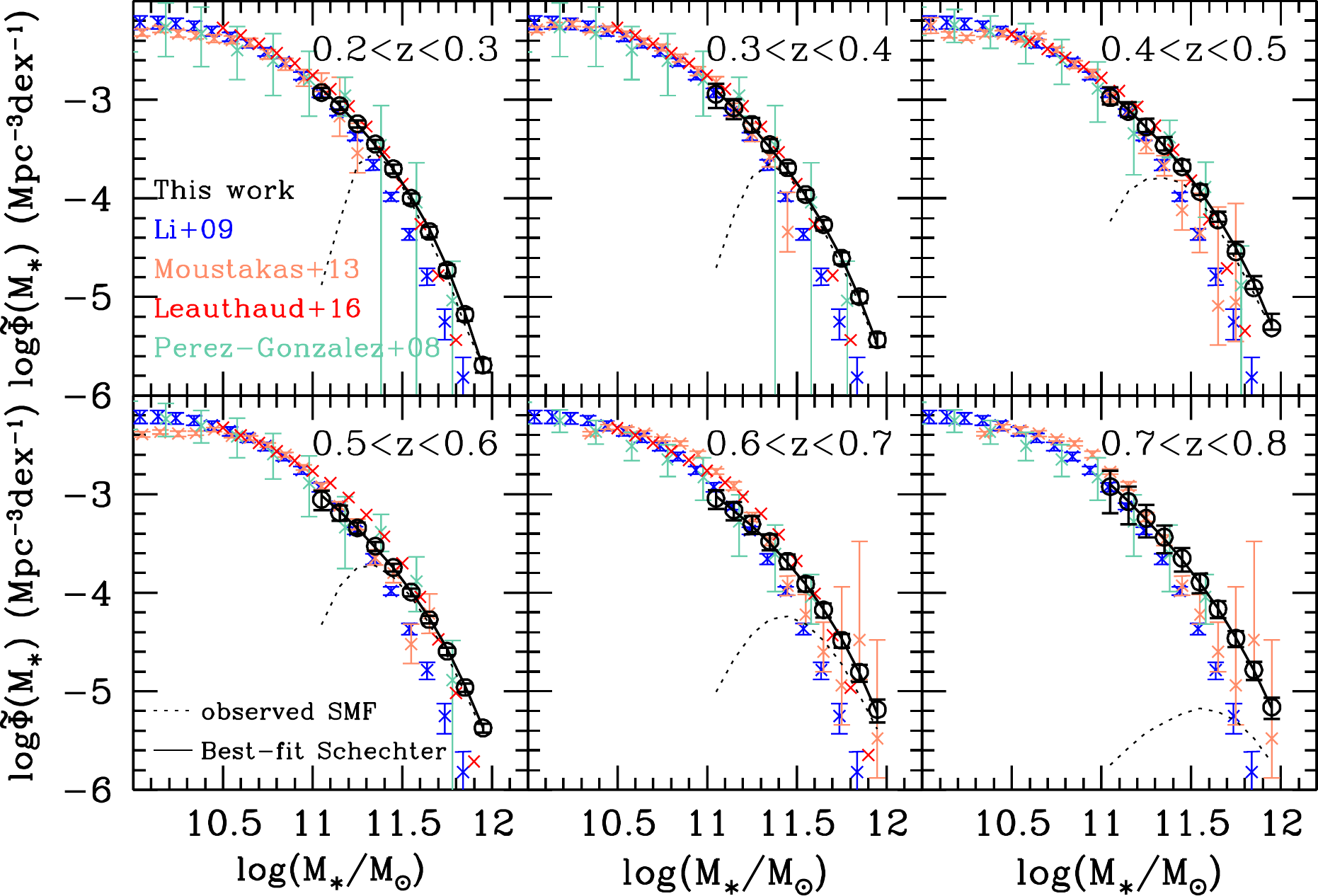}
	\caption{Total galaxy stellar mass functions,
		$\tilde{\Phi}(M_*)$, in different redshift bins. The black
		circles are our best-fitting models, while crosses of different
		colors represent the previous measurements of
		\cite{Perez-Gonzalez08}, \cite{Moustakas13} and \cite{Leauthaud16}. 
		We also show the measurements of \cite{Li09} at $z\sim0.1$ in each panel for comparison.
		We have corrected
		for the systematic offsets in the stellar mass estimates of
		\cite{Perez-Gonzalez08} to be consistent with our measurements (see
		text). We note that the measurements of \cite{Perez-Gonzalez08} are
		in bins of $\Delta z=0.2$, so they are repeated in some neighboring
		bins. The measurements of \cite{Moustakas13} at $z>0.5$ are in
		redshift bins of $0.5<z<0.65$ and $0.65<z<0.8$. So we repeat their
		measurements of $0.65<z<0.8$ for comparisons with our measurements
		at $z>0.6$. The dotted and solid lines in each panel show the observed BOSS galaxy SMF and the best-fit Schechter function to the total SMF, respectively.}
	\label{fig:smft}
\end{figure*}
\begin{table*}
	\caption{Predicted Galaxy Total Stellar Mass Function} \label{tab:smft}
	\centering
	\begin{tabular}{lrrrrrrr}
		\hline
		$\log M_*$ & $0.1<z<0.2$ & $0.2<z<0.3$ & $0.3<z<0.4$ & $0.4<z<0.5$ & $0.5<z<0.6$ & 
		$0.6<z<0.7$ & $0.7<z<0.8$ \\
	    \hline
		$11.05$ & $126.74\pm25.14$ & $118.66\pm14.11$ & $113.83\pm31.19$ & $105.17\pm21.82$ & $88.33\pm19.54$ & $90.58\pm19.90$ & $118.94\pm55.09$ \\
		
		$11.15$ & $87.78\pm13.93$ & $87.23\pm8.43$ & $82.83\pm18.89$ & $76.16\pm13.97$ & $65.19\pm11.49$ & $68.67\pm14.17$ & $84.21\pm34.97$ \\
		
		$11.25$ & $53.95\pm6.89$ & $57.15\pm4.76$ & $56.39\pm9.29$ & $52.94\pm8.51$ & $45.35\pm5.97$ & $49.68\pm10.37$ & $56.86\pm20.52$ \\
		
		$11.35$ & $29.46\pm3.42$ & $35.58\pm3.47$ & $35.01\pm4.69$ & $34.40\pm5.40$ & $29.62\pm2.99$ & $33.07\pm6.20$ & $36.63\pm11.51$ \\
		
		$11.45$ & $14.11\pm1.69$ & $19.94\pm2.08$ & $20.40\pm2.24$ & $20.82\pm2.81$ & $18.00\pm1.53$ & $20.90\pm3.60$ & $22.35\pm5.98$ \\
		
		$11.55$ & $5.84\pm0.75$ & $10.03\pm1.11$ & $10.94\pm1.18$ & $11.69\pm1.47$ & $10.16\pm0.91$ & $12.30\pm2.03$ & $12.77\pm2.86$ \\
		
		$11.65$ & $2.04\pm0.28$ & $4.61\pm0.55$ & $5.41\pm0.66$ & $6.06\pm0.77$ & $5.34\pm0.47$ & $6.69\pm1.10$ & $6.86\pm1.34$ \\
		
		$11.75$ & $0.60\pm0.09$ & $1.87\pm0.24$ & $2.48\pm0.34$ & $2.86\pm0.42$ & $2.54\pm0.23$ & $3.30\pm0.51$ & $3.45\pm0.64$ \\
		
		$11.85$ & $0.15\pm0.03$ & $0.67\pm0.09$ & $1.01\pm0.14$ & $1.23\pm0.21$ & $1.09\pm0.10$ & $1.56\pm0.31$ & $1.65\pm0.34$\\
		
		$11.95$ & $0.03\pm0.01$ & $0.20\pm0.03$ & $0.37\pm0.06$ & $0.48\pm0.10$ & $0.42\pm0.04$ & $0.65\pm0.17$ & $0.69\pm0.16$  \\							
		\hline
	\end{tabular}
	 
	\medskip
	The stellar mass is in units of $\msun$. The stellar mass
        function measurements are in units of
        $10^{-5}\rm{Mpc}^{-3}\rm{dex}^{-1}$. 
\end{table*}

\subsection{Total Galaxy Stellar Mass Functions}

As we demonstrated in Figure~\ref{fig:mock_compcs}, with the accurate
modeling of the observed galaxy SMFs and 2PCFs, we are able to make
accurate predictions of the total galaxy SMFs as well.  Taking
advantage of the large galaxy samples of BOSS, we show with black
circles in Figure~\ref{fig:smft} the total SMFs for galaxies in
different redshift bins. For reference, the related values are listed
in Table~\ref{tab:smft}. The measurement errors have included the
fractional errors of the observed galaxy SMFs in Figure~\ref{fig:smf},
which are added in quadrature to the model uncertainties from the MCMC
chains. For comparison, we also display the galaxy SMFs obtained by
\cite{Perez-Gonzalez08}, \cite{Moustakas13} and \cite{Leauthaud16} in similar redshift
ranges. The measurements of \cite{Li09} at $z\sim0.1$ are shown in each panel for easy comparison of the SMF evolution. The observed (uncorrected) BOSS galaxy SMF $\Phi(M_*)$ is displayed as the dotted line in each panel. The measurements of \cite{Perez-Gonzalez08} were obtained with
the IMF of \cite{Salpeter55}, the PEGASE SPS model \citep{Fioc97}, and
the dust model of \cite{Charlot00}. As suggested by \cite{Bernardi10}
(their Table~2), we apply a correction of $-0.22$~dex ($-0.25$~dex for
IMF and 0.03~dex for SPS model) to their stellar masses in order to be
consistent with our measurements \citep[see
also][]{Rodriguez-Puebla17}. We note that \cite{Rodriguez-Torres16} also presented the expected total SMF for CMASS galaxies at $0.55<z<0.65$ (their Figure~3). It was constructed by combining the observed CMASS SMF for $M_*>2.5\times10^{11}\msun$ and the SMF from \cite{Guo10} for lower masses. Since we are using the same set of CMASS galaxies, we do not show it here.

The measurements of \cite{Perez-Gonzalez08} (surveyed area $\sim$
0.184~$\deg^2$), \cite{Moustakas13} (surveyed area $\sim$
9~$\deg^2$) and \cite{Leauthaud16} (surveyed area $\sim$ 139.4~$\deg^2$) are limited by the small survey volumes and suffer from
significant sample variance effects. From the comparisons, our SMF
measurements are basically consistent with the trend seen in the
previous ones, but we present the most accurate measurements for
massive galaxies of $M_*>10^{11}\msun$. At $z>0.6$ where significant
differences between the SMF measurements appear in literature, our
measurements tend to agree better with those of
\cite{Perez-Gonzalez08}.  Since the galaxy SMF measurements are
generally used to infer the galaxy SHMR, it is therefore important to
use the accurate galaxy SMFs to derive the galaxy SHMR at the massive
end.  Finally, we note that as the overall completeness for $z>0.7$
galaxy sample is very low in BOSS, we expect future larger surveys
will obtain more accurate measurements of the galaxy total SMFs at
these higher redshifts.

To ease comparisons with literature, we fit our measurements of the galaxy total SMF with a standard single Schechter function \citep{Schechter76},
\begin{equation}
\Phi(M_*)=(\ln 10)\Phi^{\ast}\exp\left(-\frac{M_*}{M^{\ast}_{\rm c}}\right)\left(\frac{M_*}{M^{\ast}_{\rm c}}\right)^{1+\alpha^\ast}
\end{equation}
The best-fit Schechter functions for the galaxy samples are shown as the solid lines in Figure~\ref{fig:smft}. The best-fit parameters $\Phi^{\ast}$, $M^{\ast}_{\rm c}$ and $\alpha^\ast$ are displayed in Table~\ref{tab:schechter}. The low-mass end slope $\alpha^\ast$ is not well constrained by our measurements as we are only using the massive end of the SMF.

\begin{figure*}
	\centering
	\includegraphics[width=0.8\textwidth]{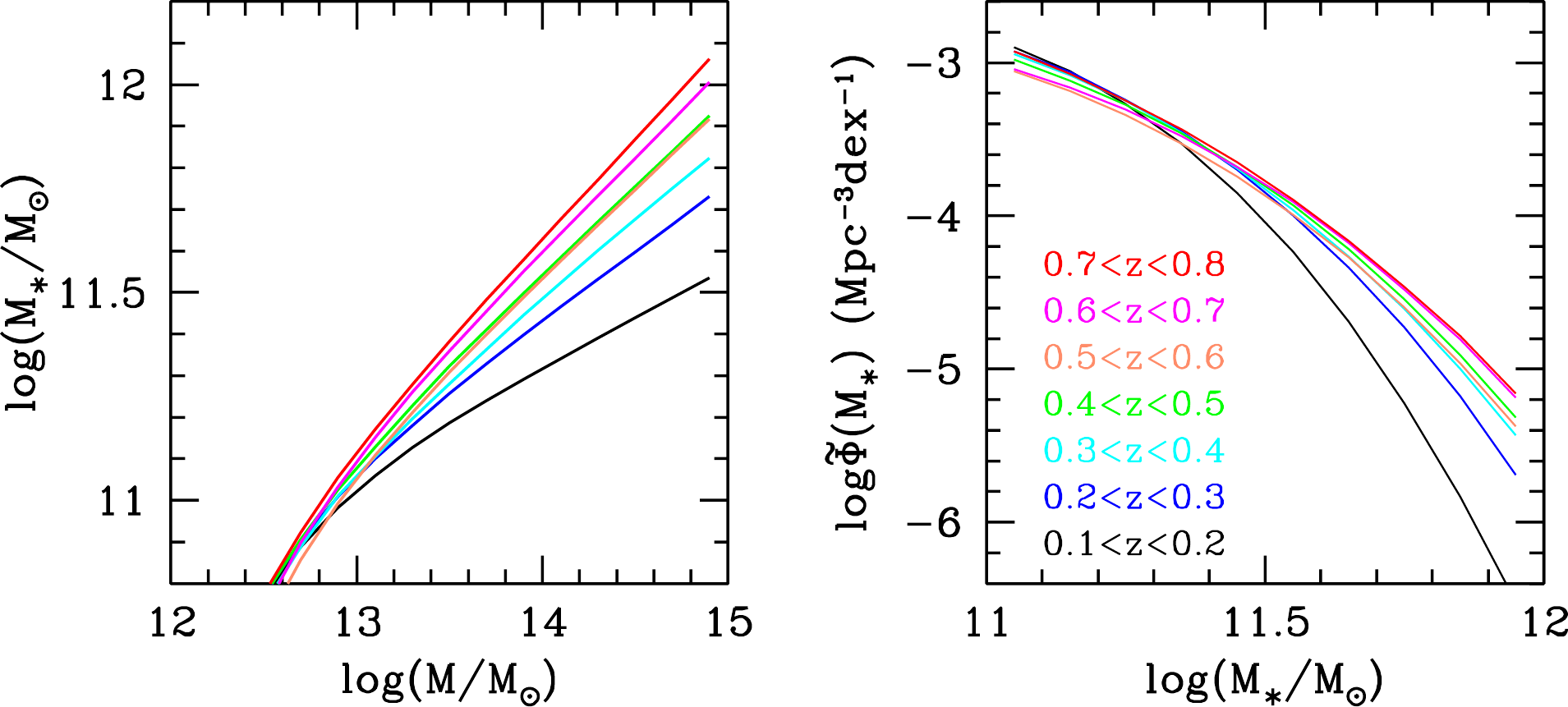}
	\caption{Redshift evolution of the galaxy SHMR (left) and total SMF
		(right). The best-fitting models at different redshifts are shown
		using lines of different colors.}
	\label{fig:evol}
\end{figure*}

\begin{table}
	\caption{Best-fit Schechter Function Parameters} \label{tab:schechter}
	\centering
	\begin{tabular}{cccc}
		\hline
		$z$ range & $\log\Phi^\ast$ & $\log M^\ast_{\rm c}$ & $\alpha^\ast$  \\
		\hline
		$0.1<z<0.2$ & $16.27\pm2.76$ & $11.05\pm0.05$ & $-1.87\pm0.34$ \\
		
		$0.2<z<0.3$ & $6.94\pm1.43$ & $11.24\pm0.04$ & $-1.97\pm0.21$ \\
		
		$0.3<z<0.4$ & $3.70\pm1.43$ & $11.36\pm0.07$ & $-2.22\pm0.29$ \\
		
		$0.4<z<0.5$ & $2.99\pm1.33$ & $11.41\pm0.08$ & $-2.15\pm0.27$ \\
		
		$0.5<z<0.6$ & $2.84\pm0.79$ & $11.39\pm0.06$ & $-2.09\pm0.22$ \\
		
		$0.6<z<0.7$ & $3.00\pm1.60$ & $11.40\pm0.11$ & $-1.95\pm0.35$ \\
		
		$0.7<z<0.8$ & $1.95\pm1.91$ & $11.50\pm0.17$ & $-2.34\pm0.51$ \\						
		\hline
	\end{tabular}
	
	\medskip
	The parameters $\Phi^\ast$ and $M^{\ast}_{\rm c}$ are in units of
	$10^{-4}\rm{Mpc}^{-3}\rm{dex}^{-1}$ and $\msun$, respectively.  
\end{table}

\subsection{Redshift Evolution}

Once we modeled the ICSMFs for different redshift samples, we are able
to study the redshift evolution of the galaxy SHMRs and total SMFs,
which are shown in the left and right panels of Figure~\ref{fig:evol},
respectively. It is clear that the slope of the galaxy SHMR is
becoming steeper at higher redshifts, while that of the total SMF is
correspondingly shallower. The general trend is consistent with the
model of \cite{Yang12} (see their Figure~13), but we have more
accurate measurements. We also note that there is only weak evolution
in the galaxy SHMR and total SMF from $z=0.4$ to $0.6$. By fitting a standard Schechter function to the total galaxy SMFs, we find that the characteristic stellar mass only varies from $10^{11.39}\msun$ to $10^{11.41}\msun$ at $0.4<z<0.6$, which supports
the conclusion of \cite{Guo13} that these massive galaxies are
consistent with passive evolution at $0.4<z<0.6$ \citep[see also][]{Tojeiro12}. Similar conclusions have been reached by \cite{Maraston13}, \cite{Montero16} and \cite{Bundy17} with the observed galaxy SMFs and luminosity functions. However, a recent
study by \cite{Bernardi16} suggests that for stellar masses obtained
using the spectral energy distribution fittings, the different light
profile fits would significant affect the abundance of the massive
galaxies. For stellar masses estimated from the PCA analysis of the
galaxy spectra, it is less affected by such an effect. But more
careful comparisons of the different methods of the stellar mass
estimates are necessary to fully resolve all the systematics, which is
beyond the scope of the current paper.

\section{Discussions} \label{sec:discussion}
An important assumption of the completeness function is that for
a given stellar mass, the selected galaxies are a random subset of the
complete sample. Here we model the central and satellite galaxies
separately, but not specifically for the color of galaxies.  The
targeting strategy of BOSS tends to select more luminous red galaxies
than blue ones \citep[see e.g., Figure~1 of][]{Guo13}. Although
galaxies with the same stellar mass but different colors are found to
occupy halos of different masses at low redshifts of $z<0.2$
\citep[see e.g.,][]{More11,More13,Paranjape15,Rodriguez-Puebla15,
	Mandelbaum16,Zu16}, the differences in the average halo masses are
estimated to be around 0.1--0.2~dex at higher redshifts of $z>0.2$ \citep[see
e.g., the right panel of Figure~7 in][]{Tinker13}. 

In this paper, we are using the whole BOSS galaxy sample to derive the completeness as a function of the stellar mass, which includes the contributions from different populations. Although we estimate the stellar mass incompleteness with the function of Eq.~\ref{eq:incomplete}, there would be possibly remaining effects of the target selections coming from the mix of the different populations of red and blue galaxies, which potentially have different selection functions.  As estimated in \cite{Masters11} and \cite{Montero16}, there are about $25\%$ of blue galaxies in the CMASS sample, while the majority of LOWZ galaxies are LRGs. \cite{Masters11} proposed to use the color cut of $g-i>2.35$ to select the red (elliptical) galaxies in CMASS \citep[see also][]{Maraston13}. In order to test whether our derived galaxy total SMF and the SHMR will be affected by the existence of blue galaxies, we impose the color cut of $g-i>2.35$ on the typical BOSS sample at $0.5<z<0.6$ to select a roughly homogeneous population of red galaxies. The fraction of blue galaxies removed in this sample is about $21\%$. Due to the existence of the photometric errors, such a simple color cut does not ensure that we have a purely red galaxy sample, but it still serves as a simple test of the effect of the blue population.

\begin{figure*}
\centering
\includegraphics[width=0.8\textwidth]{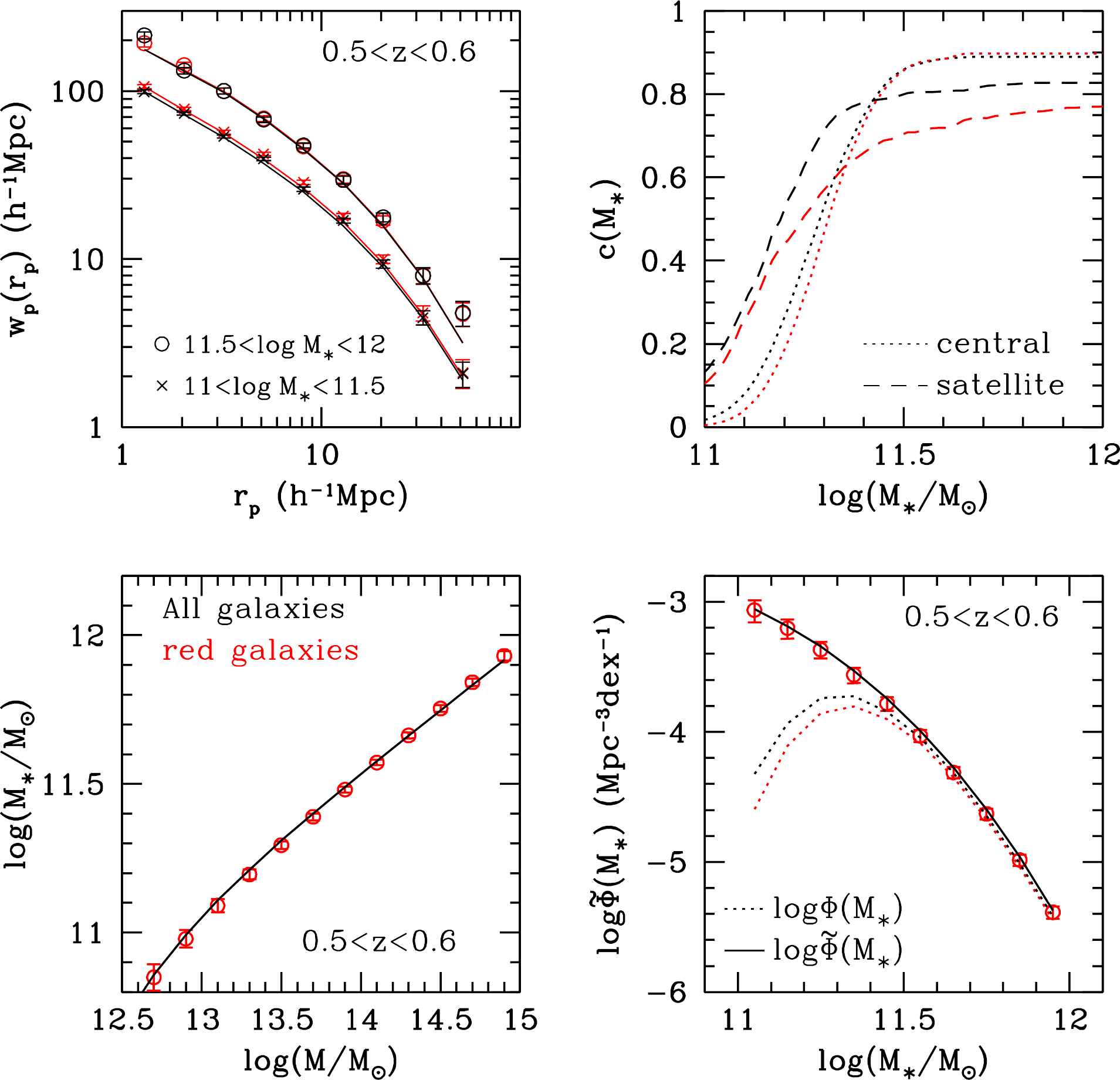}
\caption{Similar to Figure~\ref{fig:sdss}, but for the comparisons
  between the measurements and models for all galaxies (our fiducial
  model, shown as black symbols and lines) and those of red galaxies
  (red symbols and lines). We show the completeness functions of
  central and satellite galaxies in the top right panel as the dotted
  and dashed lines, respectively. The SHMR of the red galaxies is
  displayed as the red circles in the bottom left panel, while that of
  all galaxies is shown as the solid line. The observed SMFs of all
  and red galaxies are shown as the black and red dotted lines in the
  bottom right panel, respectively. The predicted total SMF of the red
  galaxies is shown as the circles and that of all galaxies is
  represented by the solid line.}
\label{fig:color}
\end{figure*}

We show in Figure~\ref{fig:color} the comparisons
  between the measurements and models for all galaxies (our fiducial
  model, shown as black symbols and lines) and those of red galaxies
  (red symbols and lines). As the red galaxies dominate the SMF at
  $M_*>10^{11}\msun$, the observed clustering measurements of $\wprp$
  are only slightly affected by the existence of blue galaxies for the
  lower stellar mass bin, while the derived SHMR and the total galaxy
  SMF are quite consistent with those obtained from all BOSS
  galaxies.

The most relavent change for our red galaxy sample is
  in the stellar mass completeness shown in the top right panel of
  Figure~\ref{fig:color}, where the completeness functions of central
  and satellite galaxies are displayed as the dotted and dashed lines,
  respectively. Although there seems to be larger decreases of
  satellite completeness at the massive end, this does not imply that
  we removed more blue satellite galaxies. Note that at the very
  massive end, according to comparison between the observed SMFs of
  all and red galaxies in the bottom right panel, there are almost no
  blue galaxies at all. Then at the relatively low mass end with
  $M_*<10^{11.3}\msun$, both the central and satellite galaxy
  completeness functions are slightly reduced by a roughly similar
  amount (or slightly larger for the satellite galaxies) due to the
  removal of blue galaxies.  However, considering that the satellite
  fraction is only about 10\% \citep{White11,Guo14} at this mass
  range, the majority of the removed blue galaxies should still be
  central galaxies.  Indeed, the fractional decrease of the observed
  SMF at the low mass end for the red galaxies is consistent with the
  decrease of amplitude for the central galaxy completeness.

The above test demonstrates that even though we are
  using the whole BOSS sample to derive the SHMR and the total galaxy
  SMF for $M_*>10^{11}\msun$, our results are not affected by the
  existence of blue galaxies. It is caused by the fact that the red
  galaxies dominate the massive end of the SMF. As discussed in
  \cite{Tinker13}, the possible differences between the SHMRs of red
  and blue galaxies are not large enough to significantly change the
  clustering for these massive galaxies. However, we note that if one
  goes to much lower mass galaxies where red and blue galaxy
  populations may have quite different stellar to halo mass relations,
  one may include a separate component to represent the fractions of
  red and blue galaxies, respectively. This separation is beyond the
  scope of current paper and will be explored in a future work.

Considering the construction of the theoretical model, we assumed a
double power-law functional form for the central galaxy SHMR and the
galaxy selection function is characterized by Eq.~\ref{eq:smhm}. A more
flexible functional form for the SHMR has been proposed by
\cite{Behroozi10} with a five-parameter model. But as tested in
\cite{Behroozi13}, compared to the four-parameter double power-law
model, the additional free parameter mainly helps improve the fits to
the SMF at the low mass end at $M_*<10^9\msun$. Adopting the
five-parameter SHMR model has minor effects on our results, as we are
focusing on the most massive galaxies. From the goodness of our fits
to the clustering measurements and observed SMFs in
Figures~\ref{fig:wp} and~\ref{fig:smf}, we conclude that the
functional form for the stellar mass completeness is flexible
enough for modeling the completeness of the BOSS galaxies, which has already been shown in \cite{Leauthaud16} and \cite{Montero16} for the completeness as a function of both stellar mass and magnitude, respectively.

Lastly, we assumed the scatter $\sigma_{\rm c}$ between the stellar
and halo masses, which includes both the contribution from the
intrinsic scatter and the statistical errors on the stellar mass
estimates, to be a constant value $\sigma_{\rm c}=0.173$.  This value
is set according to the stellar mass distribution of central galaxies
in groups or halos of given masses \citep[][]{Yang09}.  We have
tested that, if we set the amount of the total scatter to be larger or
smaller than this value (e.g. with $\sigma_c=0.2$ and 0.15,
respectively), the SHMRs thus constrained at high mass end will be
somewhat lower or larger. However, the SMFs recovered are quite
independent of this $\sigma_{\rm c}$ value, i.e., roughly consistent
within $1-\sigma$ errors.  In addition, although we do not have strong
constraints on the value of $\sigma_{\rm c}$, the inferred $\chi^2$ of
different redshift samples tend to favor our fiducial model of
$\sigma_{\rm c}=0.173$ for the BOSS galaxies. Thus we expect that the
total galaxy SMFs obtained in this work are robust. We note that the assumed scatter does not include the uncertainties in the systematic effects in the stellar mass estimates caused by the different IMFs, SPS models and dust attenuation laws, which can be reasonably corrected for using constant offsets \citep[see e.g.,][]{Moustakas13,Muzzin13,Rodriguez-Puebla17}.

\section{Conclusions} \label{sec:conclusion}

In this paper, we have introduced an incomplete conditional stellar
mass function (ICSMF) model which is applicable to large-scale galaxy
surveys with complicated target selections. By assuming suitable
functional forms for the stellar mass completeness function and the
galaxy SHMR, we are able to predict the observed galaxy clustering
measurements and the incomplete galaxy SMFs, and vice versa,
constraining the ICSMF model parameters using these observational
measurements. We tested our method using mock galaxy catalogs and then
apply it to the BOSS galaxy survey in the redshift range of
$0.1<z<0.8$. We then predicted the galaxy total SMF measurements and
central galaxy SHMRs for $10^{11}\msun<M_*<10^{12}\msun$, which are
useful for studying the star formation history and evolution of these
massive galaxies.

Our main conclusions are summarized as follows.

\begin{itemize}
	\item Based on tests using mock galaxy catalogs, we show that the
	ICSMF model can accurately recover the incompleteness factors, the
	SHMRs and the galaxy total stellar mass functions. The
	incompleteness factors thus obtained is independent and more
	consistent than the methods of involving external measurements from
	other surveys, which might introduce additional systematics from
	different surveys.
	
	\item By applying our ICSMF model to the BOSS galaxy samples, we find
	that the BOSS galaxies are more than 80\% complete at the massive
	end for $0.1<z<0.5$, while for higher redshifts the completeness
	decreases very fast to around 30\% at $z\sim0.75$.
	
	\item We obtain accurate measurements of the central galaxy SHMRs for
	the BOSS galaxies with $M_*>10^{11}\msun$ from $z=0.1$ to
	$z=0.8$. We find that the high-mass end slope of the SHMR is
	generally steeper than the previous measurements in literature,
	varying from about 1/4 to 1/2 in $0.1<z<0.8$.
	
	\item We provide accurate measurements of the total galaxy SMFs for
	BOSS galaxies within mass range $10^{11}\msun<M_*<10^{12}\msun$ and
	redshift range $0.1<z<0.8$, which will provide tighter constraints
	to the evolution of these massive galaxies.
\end{itemize}

\section*{Acknowledgements}

This work is supported by the National Key Basic Research Program of
China (Nos. 2015CB857002, 2015CB857003), national science foundation
of China (Nos. 11233005, 11621303, 11655002, 11773049).
H.G. acknowledges the support of the 100 Talents Program of the
Chinese Academy of Sciences. This work is also supported by a grant
from Science and Technology Commission of Shanghai Municipality
(Grants No. 16DZ2260200).

We thank the anonymous referee for the helpful comments that significantly improve the presentation of this paper.
We thank H.~J. Mo and Zheng Zheng for helpful discussions. We gratefully acknowledge the use of the High Performance Computing
Resource in the Core Facility for Advanced Research Computing at
Shanghai Astronomical Observatory. We acknowledge the Gauss Centre for
Supercomputing e.V. (www.gauss-centre.eu) and the Partnership for
Advanced Supercomputing in Europe (PRACE, www.prace-ri.eu) for funding
the MultiDark simulation project by providing computing time on the
GCS Supercomputer SuperMUC at Leibniz Supercomputing Centre (LRZ,
www.lrz.de).

\end{document}